\documentclass[a4paper,12pt]{article}
\usepackage[utf8]{inputenc}
\usepackage{multicol}
\usepackage{amsmath}
\usepackage{amssymb}

\usepackage{graphicx}
\usepackage[dvipsnames]{xcolor}
\usepackage{graphicx,times}             

\DeclareUnicodeCharacter{2212}{\textendash}

\usepackage{graphicx}
        \usepackage{dcolumn}
        \usepackage{bm}
        \usepackage[final]{hyperref}

        \usepackage{multirow}
        \usepackage[dvipsnames]{xcolor}
        \usepackage{comment}

\title{Stability of charged scalar hair on Reissner-Nordstr\"om black holes}

        \author{Muhammed Shafeeque\footnote{m.shafeeque@iitg.ac.in} \  and Malay K. Nandy\footnote{mknandy@iitg.ac.in}\\
        \small {\it Department of Physics, Indian Institute of Technology Guwahati,}\\ \small {\it Guwahati 781 039, India.}}

        \date{June 7, 2024}

\begin{document}
\maketitle

\begin{abstract}
The Israel-Carter theorem (also known as the  ``no-hair theorem'') puts a restriction on the existence of parameters other than mass, electric charge, and angular momentum of a black hole. In this context, Bekenstein proposed no-hair theorems in various black hole models with neutral and electrically charged scalar fields.

In this paper, we take the Einstein-Maxwell-charged scalar model with an electrically charged scalar field gauge-coupled to the Maxwell field surrounding a charged black hole with a static spherically symmetric metric. In particular, we consider a quadratic scalar potential without any higher order terms and we do not impose any restriction on the magnitude of  the scalar charge with respect to  the black hole charge.

With this setting, we ascertain the validity of all energy conditions coupled with the causality condition, suggesting the possibility of existence of charged hairy solutions. Consequently, we obtain, by exact numerical integration, detailed solutions of the field equations that incorporate backreaction on the spacetime due to the presence of the charged scalar field. The solutions exhibit damped oscillatory behaviours for the charged scalar hair. We also find that the electric potential is a monotonic function of the radial coordinate, as required by electrodynamics.

In order to ascertain the existence of our charged hairy solutions, we carry out dynamic stability analyses against time-dependant perturbations about the static solutions. For a definite conclusion, we employ two different methodologies. The first methodology involves a Sturm-Liouville equation, whereas the second methodology employs a Schr\"odinger-like equation, for the dynamic perturbations. We find that our solutions are stable against time-dependant perturbations by both methodologies, confirming the existence of the charged hairy solutions.
\end{abstract}


\section{Introduction}\label{sec_intro}

The Israel-Carter conjecture, or the ``no-hair theorem''\ \cite{NH1_israel67, NH2_Israel1968, NH3_carter}, asserts the uniqueness of the Schwarzschild, Reissner-Nordstr\"om, and Kerr black hole solutions. As a consequence, a black hole can be parametrised only in terms of its mass, electric charge and angular momentum. Early studies in relation to the no-hair theorem in different contexts were made by Bekenstein in the early Seventies \cite{Bekenstein_1972_1, Bekenstein_1972_2}.

Bekenstein showed in a later work \cite{Bekenstein_1995} that a minimally coupled multi-component scalar field with positive energy density cannot exist as a hairy solution which follows from conservation of the energy-momentum tensor. He further argued that a Higgs field with a potential having two or more wells cannot exist as hairy solution for uncharged black holes.

Ever since the no-hair conjecture was made, there have been searches for other possible parameters apart from mass, electric charge and angular momentum in the black hole solution.

One of the foremost studies in this direction was made by Bekenstein in 1974 \cite{BEKENSTEIN1974}, which was further extended in Ref. \cite{BEKENSTEIN1975}. Bekenstein considered an electrically charged black hole with a massless conformal scalar field non-minimally coupled to gravity via conformal coupling, $\xi R\phi^2$. This study established a scalar charge, like other parameters allowed by the no-hair theorem, may also be an admissible parameter in black hole solutions. Bekenstein's hairy solution \cite{BEKENSTEIN1974,BEKENSTEIN1975} was analysed for stability by Bronnikov and others \cite{BRONNIKOV1978_1,Bronnikov1979_2} and found that a spherically symmetric perturbation destabilizes the scalar hair. They concluded that the scalar hair decays by evaporation due to this instability, leaving no scalar hair around the black hole.

Bekenstein's black hole models \cite{BEKENSTEIN1974,BEKENSTEIN1975} with conformal scalar hair gave rise to a spate of research studies with modifications. Models with conformally coupled scalar fields that are massless \cite{martinez_zaneli_1996,Radu_Winstanley_2005}, massive \cite{Winstanley_2003,Winstanley_2005}, and self-interacting \cite{MTZ,MTZ2,martinez_troncoso_2006}, and with a cosmological constant, were considered to obtain hairy solutions.

Furthermore, Poisson \cite{Poisson_1991} examined the black hole solution in Einstein-Gauss-Bonnet gravity in $4+n$ dimensions. Assuming the $n$-dimensional internal space to be slightly perturbed from constant radius, it was found that the internal radius behaves like a scalar hair. Subsequent works on scalar hair in Einstein-Gauss-Bonnet gravity can be found in Refs. \cite{Kanti_etal_1996, Alexeyev_etal_2009, Yagi_etal_2012, Ayzenberg_etal_2014, Blazquez_Salcedo_etal_2016, Bhattacharya_Chakraborty_2017, Doneva_etal_2018, Tattersall_etal_2018, Brihaye_Hartmann_2018, Brihaye_Hartmann_2019, BRIHAYE_etal_2019, BumHoon_etal_2019}. Black hole solutions with axionic hair were studied by Campbell et al. \cite{CAMPBELL1990_1,CAMPBELL1991_dyonbh}, Lee and Weinberg \cite{Kimyeong_Weinberg_1991_axion}, and Bardoux et al. \cite{Bardoux_2012}.

There also exists a new family of Einstein-Maxwell-scalar field models where an electrically neutral scalar field $\varphi$ is minimally coupled to gravity and non-minimally coupled to the Maxwell field via the interaction $f(\varphi)F^{\mu\nu}F_{\mu\nu}$ \cite{scalar1_PhysRevLett.121.101102}. Such models undergo spontaneous scalarization when the effective mass-squared given by $f^{\prime\prime}(0)F^{\mu\nu}F_{\mu\nu}$ changes sign from positive to negative, creating (uncharged) hairy solutions by dynamical growth of the scalar field profile. A comparative study on the choice for $f(\varphi)$ is reported in Ref. \cite{Fernandes_2019} and its stability is analysed in Ref. \cite{Myung2019}. Extension of such scalarised black hole models with massless \cite{Jhep_main2019, scalar2_Fernandes_2019,scalar3, scalar4_PhysRevD}, massive \cite{massive_scalar_PhysRevD.100.124055}, massive-self-interacting \cite{massive_self_int_scalar}  scalar fields, and with negative cosmological constant \cite{Promsirietal} , have also been studied.

 In an important work, Mayo and Bekenstein \cite{Mayo_Bekenstein_1996} considered different combinations of electrically neutral as well as charged black holes with self-interacting neutral and charged scalar fields conformally coupled with gravity via $\xi R |\phi|^2$. The charged scalar field was gauge coupled to the Maxwell field. In the case of electrically neutral scalar field, they established a no-hair theorem for the conformal parameter in the ranges $\xi<0$ and $\xi\ge\frac{1}{2}$.  They however indicated that the total {\em integrated} charge should be unbounded for any value of $\xi$ in the case of charged scalar field. In order to circumvent this situation, they further made the choice of vanishing electric potential $V(r)$ at infinity. However, they found that this latter choice contradicts the fact that $V(r)$ must be a monotonic function of $r$.

 The no scalar hair theorem by Mayo and Bekenstein \cite{Mayo_Bekenstein_1996} was further examined for charged scalar fields gauge coupled with Maxwell field by Hong et al. \cite{Hong_suzuki_yamada_PLB, Hong_suzuki_yamada_PRL}. It was pointed out that the contribution of the scalar mass in the asymptotic solution was not properly taken into consideration in the analysis of Mayo and Bekenstein. They considered a charged scalar field gauge coupled to the Maxwell field with polynomial interacting potential with scalar charge being very small compared to the charge of the black hole. Such a system with a charged scalar cloud around the black hole was studied without any backreaction to the Reissner-Nordstr\"om background in Ref. \cite{Hong_suzuki_yamada_PLB} and with backreaction in Ref.  \cite{Hong_suzuki_yamada_PRL}. The latter model was further studied by Herdeiro and Radu \cite{Herdeiro_radu_2020} for Schwarzschild and Reissner-Nordstr\"om black holes with charged clouds as hair. They examined various conditions for the existence of charged scalar hair.

A common feature in these models was to take the scalar potential to be a polynomial of sixth degree and the scalar charge to be much smaller than the black hole charge. Consequently, we may ask the question whether a charged scalar hair would exist as a stable solution if the scalar field is taken to be a charged massive Klein-Gordon field (with quadratic potential) without any limitation to its charge compared to the black hole charge.

In this paper, we therefore consider a Einstein-Maxwell-charged scalar field model where an electrically charged and massive Klein-Gordon field (with quadratic potential) is gauge-coupled to the Maxwell field giving rise to an interaction of the electric charge $q$ of the scalar field with the electromagnetic field surrounding a charged black hole. Importantly, we take account of the backreaction on the exterior spacetime due to the presence of the scalar field by considering all field equations without making any approximation. We solve this model in a static spherically symmetric spacetime by numerical integration and obtain black hole solutions with electrically charged scalar hair. We also consider all energy conditions stemming from the energy-momentum tensor as well as the Poynting vector giving the causality condition, and find that all these conditions are satisfied.

We further ascertain the stability of the charged hairy solutions by carrying out dynamic stability analyses employing two different methodologies. The first methodology \cite{CS1, CS2,CS3, Bardeen_1966} involves an analysis based on the Sturm-Liouville equation for sinusoidal perturbations about the static hairy solutions. The second methodology \cite{BRONNIKOV1978_1,Bronnikov1979_2} involves an analysis of a Schr\"odinger-like equation for the dynamic perturbations. We find that the charged hairy solutions are stable against dynamical perturbations in both methodologies.

The rest of the paper is organised as follows. In Section \ref{sec_model}, we define the Einstein-Maxwell-charged scalar field model and obtain the corresponding field equations using a static spherically symmetric metric. In Section \ref{sec_Tmunu_analysis}, we ascertain the validity of all energy conditions coupled with causality condition. In Section \ref{sec_sol}, we describe the boundary conditions for the field equations and solve them exactly by numerical integration. There, we also describe the hairy solutions and illustrate them with several plots. Section \ref{sec_stability} is devoted to detailed stability analyses of the hairy solutions obtained in the previous Section where we employ two different methodologies to ascrtain the stability of the solutions. Section \ref{sec_conc} gives a discussion and conclusion of the paper.


\section{Einstein-Maxwell-charged scalar field model}\label{sec_model}
We consider the Einstein-Maxwell model gauge coupled to a charged scalar field described by the action,
\begin{equation}\label{eq_action_1}
S=\int d^{4}x\sqrt{-g}\left[\frac{R}{2}-\left(D_{\mu}\phi\right)^{*}D^{\mu}\phi-\mu^{2}\phi^{*}\phi-\frac{1}{4}F_{\mu\nu}F^{\mu\nu}\right],
\end{equation}
where $D_{\mu}$ is the covariant derivative $D_{\mu}=\partial_{\mu}+iqA_{\mu}$, with $A_\mu$ the four-potential of the Maxwell field $F_{\mu\nu}=\partial_\mu A_\nu - \partial_\nu A_\mu$, and $q$ and $\mu$ are the charge and mass of the scalar field, respectively. The gauge coupling term generates an interaction between the scalar field and the electromagnetic field.

In the above Einstein-Maxwell-charged scalar field model, we shall explore the possibility of obtaining black hole solutions with charged scalar hair for various parameter values. We shall also perform stability analyses of such hairy solutions employing two different methodologies.


We first obtain the field equations from the extremum of the above action (\ref{eq_action_1}) with respect to the variations $\delta g_{\mu\nu}$, $\delta \phi^*$, and $\delta A_\mu$. As a consequence, we obtain the modified Einstein field equation,
\begin{equation}
\begin{aligned}G_{\mu\nu}=   & 2\left(\partial_{\mu}-iqA_{\mu}\right)\phi^{*}\left(\partial_{\nu}+iqA_{\nu}\right)\phi-g_{\mu\nu}\left\{ \left(\partial_{\rho}-iqA_{\rho}\right)\phi^{*}\left(\partial^{\rho}+iqA^{\rho}\right)\phi+\mu^{2}\phi^{*}\phi\right\} \\
 & +F_{\mu}^{\ \sigma}F_{\nu\sigma}-\frac{g_{\mu\nu}}{4}F_{\rho\sigma}F^{\rho\sigma},
\end{aligned}
\end{equation}
the modified Klein-Gorden equation,
\begin{equation}
 \Box_{g}\phi+iq\phi\nabla_{\mu}A^{\mu}+2iqA^{\mu}\nabla_{\mu}\phi-q^{2}A_{\mu}A^{\mu}\phi-\mu^{2}\phi=0,
\end{equation}
and the modified Maxwell equation,
\begin{equation}
 \nabla_{\mu}F^{\mu\nu}-2q^{2}A^{\nu}\phi^{*}\phi-iq\left(\phi\partial^{\nu}\phi^{*}-\phi^{*}\partial^{\nu}\phi\right)=0.
\end{equation}

Upon writing the complex scalar field as $\phi=\frac{1}{\sqrt{2}}\left(\psi+i\chi\right)$, the above field equations can be rewritten as
\begin{equation}
\begin{aligned}G_{\mu\nu} & =\partial_{\mu}\psi\partial_{\nu}\psi+\partial_{\mu}\chi\partial_{\nu}\chi+q^{2}A_{\mu}A_{\nu}\left(\psi^{2}+\chi^{2}\right)+q\psi\left(A_{\mu}\partial_{\nu}\chi+A_{\nu}\partial_{\mu}\chi\right)\\
 & -q\chi\left(A_{\mu}\partial_{\nu}\psi+A_{\nu}\partial_{\mu}\psi\right)+F_{\mu}^{\ \sigma}F_{\nu\sigma}-\frac{g_{\mu\nu}}{4}F_{\rho\sigma}F^{\rho\sigma}\\
 & -\frac{g_{\mu\nu}}{2}\left\{ \partial_{\rho}\psi\partial^{\rho}\psi+\partial_{\rho}\chi\partial^{\rho}\chi+\left(\mu^{2}+q^{2}A_{\rho}A^{\rho}\right)\left(\psi^{2}+\chi^{2}\right)+2q\left(\psi A_{\rho}\partial^{\rho}\chi-\chi A_{\rho}\partial^{\rho}\psi\right)\right\} ,
\end{aligned}
\label{eq_einstein_real}
\end{equation}

\begin{equation}
\begin{aligned}\Box_{g}\psi-q\chi\nabla_{\mu}A^{\mu}-2qA^{\mu}\nabla_{\mu}\chi- & \left(\mu^{2}+q^{2}A_{\mu}A^{\mu}\right)\psi=0,\end{aligned}
\label{eq_KG_phi_1}
\end{equation}

\begin{equation}
\begin{aligned}\Box_{g}\chi+q\psi\nabla_{\mu}A^{\mu}+2qA^{\mu}\nabla_{\mu}\psi & -\left(\mu^{2}+q^{2}A_{\mu}A^{\mu}\right)\chi=0,\end{aligned}
\label{eq_KG_phi_2}
\end{equation}
and
\begin{equation}
\nabla_{\mu}F^{\mu\nu}-q^{2}A^{\nu}\left(\psi^{2}+\chi^{2}\right)-q\left(\psi\partial^{\nu}\chi-\chi\partial^{\nu}\psi\right)=0\label{eq_maxwell}
\end{equation}

We consider a static and spherically symmetric metric
\begin{equation}
ds^{2}=-f(r)e^{-2\eta(r)}dt^{2}+\frac{dr^{2}}{f(r)}+r^{2}\left(d\theta^{2}+\sin^{2}\theta d\varphi^2\right),
\end{equation}
and obtain the equations for the metric coefficients $f(r)$ and $\eta(r)$ employing equations \ref{eq_einstein_real}, \ref{eq_KG_phi_1}, \ref{eq_KG_phi_2} and (\ref{eq_maxwell}).

For simplicity in obtaining the solutions, we consider the ansatz $A_\mu=\left(V(r),0,0,0\right)$, where $V(r)$ is the electrostatic potential. Using this ansatz in (\ref{eq_maxwell}), we have the four-current
\begin{equation}\label{eq_J_nu}
\mathcal{J}^{\nu}=\left(\psi\partial^{\nu}\chi-\chi\partial^{\nu}\psi\right)=0.
\end{equation}
Thus, as a consequence of this simplifying ansartz, we see that $\psi$ and $\chi$ are related as $\chi=\alpha\psi$. Substituting this in (\ref{eq_KG_phi_1}) and (\ref{eq_KG_phi_2}), we find that the proportionality constant can be fixed as  $\alpha=1$.

Employing this ansatz in the Maxwell equation (\ref{eq_maxwell}) yields the differential equation for the electrostatic potential as

\begin{equation}\label{eq_fV_dd}
V''+\left(\eta'(r)+\frac{2}{r}\right)V'-\frac{2q^{2}}{f}V\psi^{2}=0,
\end{equation}
where a prime denotes differentiation with respect to $r$. Here $V^\prime(r)$ is the magnitude of the electric field due to electric charges of both the black hole as well as the scalar field enclosed by a sphere of radius $r$. In addition, from the Einstein field equation (\ref{eq_einstein_real}), we obtain

\begin{equation}\label{eq_f_and_h_d}
f'=\frac{1-f}{r}-r\left\{ f\left(\psi'\right)^{2}+\frac{e^{2\eta(r)}}{2}\left(V'\right)^{2}+\left(\mu^{2}+q^{2}\frac{e^{2\eta(r)}}{f}V^{2}\right)\psi^{2}\right\}
\end{equation}
and
\begin{equation}\label{eq_eta_d}
 \eta'(r) =-r\left[\left(\psi'\right)^{2}+q^{2}\frac{e^{2\eta(r)}}{f^{2}}V^{2}\psi^{2}\right].
\end{equation}

Similarly, equation (\ref{eq_KG_phi_1}) leads to

\begin{equation}\label{eq_KG}
\psi'' +\left(\frac{f'}{f}-\eta'+\frac{2}{r}\right)\psi' + \left(q^{2}\frac{e^{2\eta}}{f^{2}}V^{2}-\frac{\mu^{2}}{f}\right)\psi=0.
\end{equation}

The above set of field equations (\ref{eq_fV_dd}), (\ref{eq_f_and_h_d}), (\ref{eq_eta_d}) and (\ref{eq_KG}) are nonlinear in addition to being coupled in a very complicated way. Consequently, in order to obtain exact solutions, we solve them numerically with appropriate boundary conditions at the horizon. We lay out the details of the numerical procedure in Section \ref{sec_sol}.

\section{Energy conditions for charged scalar hair}\label{sec_Tmunu_analysis}
Before obtaining the exact solutions, it is important to examine all relevant energy conditions supplemented with the causality conditions in the Einstein-Maxwell-charged scalar field model.

We thus require that the energy-momentum tensor with mixed components $T^\mu_{\ \nu}$ must be bounded and it should not violate causality conditions. In order to have no divergences in the curvature invariant $G_{\mu\nu}G^{\mu\nu}$, the energy-momentum invariant $T_{\mu\nu}T^{\mu\nu}$ must be bounded everywhere in the exterior including the horizon. In the static and spherically symmetric geometry, $T_{\mu\nu}T^{\mu\nu}=(T^t_{\ t})^2+(T^r_{\ r})^2+(T^\theta_{\ \theta})^2+(T^\varphi_{\ \varphi})^2$. Thus the components $T^t_{\ t}$, $T^r_{\ r}$, and $T^\theta_{\ \theta}=T^\varphi_{\ \varphi}$ should be finite and well-behaved everywhere including the horizon.

In addition, the Poynting vector $j^\mu=-T^\mu_{\ \nu}u^\nu$ gives the causality condition $T^\mu_{\ \nu}u^\nu T^\rho_{\ \mu}u_\rho\le0$ with respect to  a distant observer with the four-velocity $u^\nu$. This condition can be simplified using $u^\mu u_\mu=-1$ along with the fact that $u^ru_r$ and $u^\varphi u_\varphi$ are always positive for any observer moving in the equatorial plane, yielding the the condition
\begin{equation}\label{eq_energy_condition}
 |T^\theta_{\ \theta}|=|T^\varphi_{\ \varphi}|\le|T^t_{\ t}|\ge|T^r_{\ r}|.
\end{equation}
These energy condition were discussed by Hawking and Ellis \cite{Hawking_Ellis_1973} and Mayo and Bekenstein \cite{Mayo_Bekenstein_1996}.
\begin{figure}[t!]
 \centering
 \includegraphics[width=.8\textwidth, height=9cm]{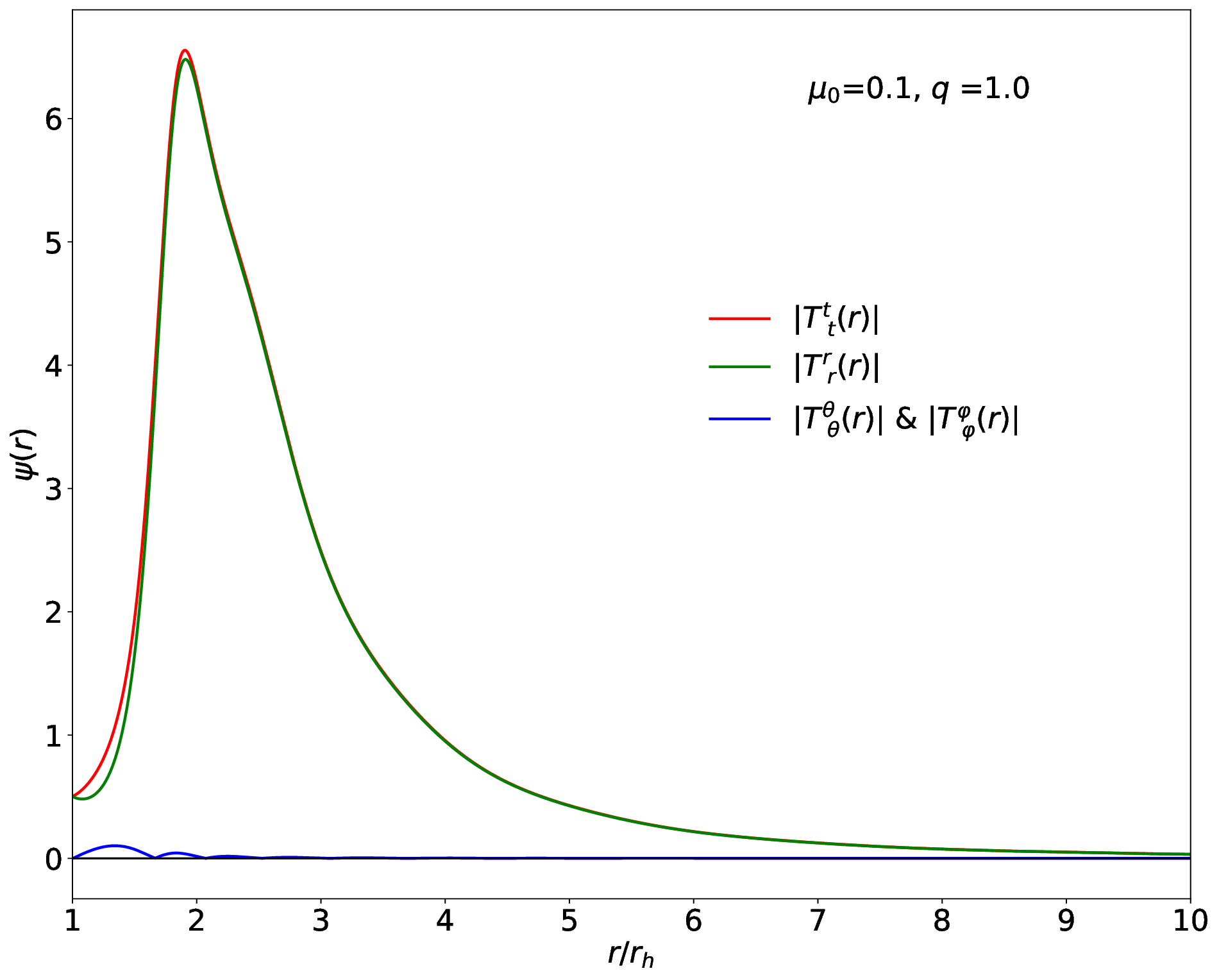}
 \caption{\footnotesize Radial profiles of the mixed components of energy-momentum tensor  $|T^t_{\ t}(r)|$, $|T^r_{\ r}(r)|$, $|T^\theta_{\ \theta}(r)|$, and $|T^\varphi_{\ \varphi}(r)|$ with $q=1.0$ and $\mu=0.1$ in units of $r_h$. The boundary conditions chosen are $V'(r_h)=1.0$ and $\psi(r_h)=0.5$.}
 \label{fig.radprof_EC}
\end{figure}

For the action given by Equation (\ref{eq_action_1}), the energy-momentum tensor can be obtained from the right-hand side of the Einstein equation (\ref{eq_einstein_real}) as
\begin{equation}\label{eq_EM_tensor}
\begin{aligned}T_{\mu\nu} & =\partial_{\mu}\psi\partial_{\nu}\psi+\partial_{\mu}\chi\partial_{\nu}\chi+q^{2}A_{\mu}A_{\nu}\left(\psi^{2}+\chi^{2}\right)+q\psi\left(A_{\mu}\partial_{\nu}\chi+A_{\nu}\partial_{\mu}\chi\right)\\
 & -q\chi\left(A_{\mu}\partial_{\nu}\psi+A_{\nu}\partial_{\mu}\psi\right)+F_{\mu}^{\ \sigma}F_{\nu\sigma}-\frac{g_{\mu\nu}}{4}F_{\rho\sigma}F^{\rho\sigma}\\
 & -\frac{g_{\mu\nu}}{2}\left\{ \partial_{\rho}\psi\partial^{\rho}\psi+\partial_{\rho}\chi\partial^{\rho}\chi+\left(\mu^{2}+q^{2}A_{\rho}A^{\rho}\right)\left(\psi^{2}+\chi^{2}\right)+2q\left(\psi A_{\rho}\partial^{\rho}\chi-\chi A_{\rho}\partial^{\rho}\psi\right)\right\} ,
\end{aligned}
\end{equation}
Following the discussion in the previous Section \ref{sec_model} for a static and spherically symmetric geometry, we obtain the mixed components $T^\mu_{\  \nu}$ as
\begin{equation}\label{eq_EM_tensor_mixed}
 \begin{aligned}-T_{\ t}^{t}= & f\psi^{\prime2}+\frac{e^{2\eta}V^{\prime2}}{2}+\frac{q^{2}V^{2}}{fe^{-2\eta}}\psi^{2}+\mu^{2}\psi^{2}\\
T_{\ r}^{r}= & f\psi^{\prime2}-\frac{e^{2\eta}V^{\prime2}}{2}+\frac{q^{2}V^{2}}{fe^{-2\eta}}\psi^{2}-\mu^{2}\psi^{2}\\
T_{\ \theta}^{\theta}= & T^\varphi_{\ \varphi}=-f\psi^{\prime2}+\frac{q^{2}V^{2}}{fe^{-2\eta}}\psi^{2}-\mu^{2}\psi^{2}.
\end{aligned}
\end{equation}

As we shall see in Sec. \ref{sec_sol}, we must have the boundary condition $V(r=r_h)=0$ at the horizon for consistency with the field equations. To analyse the behaviours of the above mixed components at the horizon, we may assume the metric potential to have the Reissner-Nordstr\"om form, $f(r)=1-\frac{2M(r)}{r}+\frac{Q^2(r)}{r^2}$, and the electric potential $V(r)=\frac{Q(r)}{r}-\frac{Q_h}{r_h}$. Very close to the horizon, we may write  $r=r_h + \epsilon$, so that $f(r)=f_1\epsilon+O(\epsilon^{2})$ and $V(r)=V_1\epsilon+O(\epsilon^2)$, where $f_{1}=\frac{2M_{h}}{r_{h}^{2}}-\frac{2M_{h}^{\prime}}{r_{h}}-\frac{2Q_{h}^{2}}{r_{h}^{3}}+\frac{2Q_{h}Q_{h}^{\prime}}{r_{h}^{2}}$ and $V_{1}=\frac{Q_{h}^{\prime}}{r_{h}}-\frac{Q_{h}}{r_{h}^{2}} $, giving $\frac{V^{2}(r)}{f(r)}=\frac{V_1^2}{f_1}\epsilon+O(\epsilon^{2})$, which is zero at the horizon ($\epsilon\rightarrow0$).

Thus all terms in each mixed component given by (\ref{eq_EM_tensor_mixed}) are finite and well-behaved at the horizon for finite boundary conditions on $\psi(r=r_h)$, $\psi^\prime(r=r_h)$, together with $\eta(r=r_h)=0$. Consequently, all energy conditions, supplemented by the causality condition, given by the inequalities in (\ref{eq_energy_condition}) are readily satisfied. In order to illustrate the validity of these energy conditions, we show in Figure \ref{fig.radprof_EC} the radial profiles of the mixed components of the energy momentum tensor.

In this context, we note that no violation of these energy conditions provide only a {\em necessary} criterion for the existence of scalar hair. The {\em sufficient} condition for the existence of scalar hair is provided by the stability of a hairy solution. (On the other hand, violation of any of the above energy conditions would be sufficient to prove a no-hair theorem.)

In the next Section \ref{sec_sol}, we shall solve the coupled field equations exactly by numerical integration to obtain hairy solutions. In the subsequent Section \ref{sec_stability}, we shall analyse the stability of the hairy solutions employing two different methodologies.

\section{Exact solutions of the field equations}\label{sec_sol}
We carry out numerical integrations for the coupled set of nonlinear differential equations
(\ref{eq_fV_dd}), (\ref{eq_f_and_h_d}), (\ref{eq_eta_d}) and (\ref{eq_KG}) starting from the horizon.

\subsection{Horizon boundary conditions}
In order to set up the boundary conditions for the gradients of the fields, we expand them about the horizon, $r=r_{h}$, as

\begin{equation}\label{eq_horizon_exp}
\begin{aligned}f(r) & =f_{0}+(r-r_{h})f_{1}+\frac{1}{2}(r-r_{h})^{2}f_{2}+\ldots\\
\eta(r) & =\eta_{0}+(r-r_{h})\eta_{1}+\frac{1}{2}(r-r_{h})^{2}\eta_{2}+\ldots\\
\psi(r) & =\psi_{0}+(r-r_{h})\psi_{1}+\frac{1}{2}(r-r_{h})^{2}\psi_{2}+\ldots\\
V(r) & =V_{0}+(r-r_{h})V_{1}+\frac{1}{2}(r-r_{h})^{2}V_{2}+\ldots
\end{aligned}
\end{equation}
where  we have used the notation $Y_0=Y(r_h)$, $Y_1=Y'(r_h)$ and $Y_2=Y''(r_h)$.

We note that $f_0=0$ and we take $\eta_0=0$ without any loss of generality since $e^{-2\eta(r)}$ corresponds to a gauge choice in the definition of the time coordinate.

Substituting the expansions given by (\ref{eq_horizon_exp}) in the Maxwell equation (\ref{eq_fV_dd}), and equating powers of the same order in $r-r_h$, we find $V_0\psi_0^2=0$. For a general boundary condition $\psi_0\not=0$, we therefore must choose the boundary condition $V_0=0$ at the horizon.

We further obtain the horizon conditions by expanding the field equations about the horizon. Upon substituting $r=r_h + \epsilon$ in equations (\ref{eq_fV_dd}), (\ref{eq_f_and_h_d}), (\ref{eq_eta_d}) and (\ref{eq_KG}), the radial derivatives at the horizon are obtained as
\begin{equation}\label{eq_near_hor}
\begin{aligned}f_{1} & =\frac{1}{r_{h}}-\mu^{2}r_{h}\psi_{0}^{2}-\frac{r_{h}V_{1}^{2}}{2}\\
\psi_{1} & =\frac{\mu^{2}}{f_{1}}\psi_{0}\\
\eta_{1} & =-r_{h}\psi_{1}^{2}-q^{2}r_{h}\psi_{0}^{2}\frac{V_{1}^{2}}{f_{1}^{2}}.
\end{aligned}
\end{equation}
The parameter $\epsilon$ is a very small radial displacement from the horizon, which will be controlled in the numerical algorithm.

 On the other hand, we expect a Minkoskian spacetime as $r\rightarrow\infty$ so that $f\rightarrow 1$. Moreover, with the gauge choice of $V=0$ at the horizon, $V(r)$ and $\eta(r)$  approach constant values as $r\rightarrow\infty$. With these asymptotic behaviours, equation (\ref{eq_KG}) reduces to $\psi''+\frac{2}{r}\psi'+\omega^{2}\psi=0$ at infinity, where $\omega^{2}=q^{2}e^{2\eta_\infty}V_\infty^{2}-\mu^{2}=$ constant. The corresponding solutions that satisfy the above asymptotic requirements is $\psi\propto\frac{\cos(\omega r)}{r}$ for $\omega^2 >0$  and $\psi\propto\frac{e^{-\omega r}}{r}$ for $\omega^2 <0$. Both solutions vanish as $r\rightarrow\infty$.  However, the latter solution is not possible in the present model with boundary conditions discussed above.

\subsection{Numerical integration}

Employing the boundary conditions at the horizon as discussed in the previous subsection, we carry out simultaneous numerical integrations of the coupled set of field equations (\ref{eq_fV_dd}), (\ref{eq_f_and_h_d}), (\ref{eq_eta_d}) and (\ref{eq_KG}) by choosing the parameters $V_1$, $\psi_0$, $q$ and  $\mu$, so as to obtain exact solutions for the radial profiles of $\psi(r)$, $V(r)$, $f(r)$, and $\eta(r)$, until the radial distance from the black hole horizon is sufficiently high, where the scalar field approaches a vanishing value. The results of the numerical integration are presented below.
\begin{figure}[t!]
 \centering
 \includegraphics[width=.8\textwidth, height=9cm]{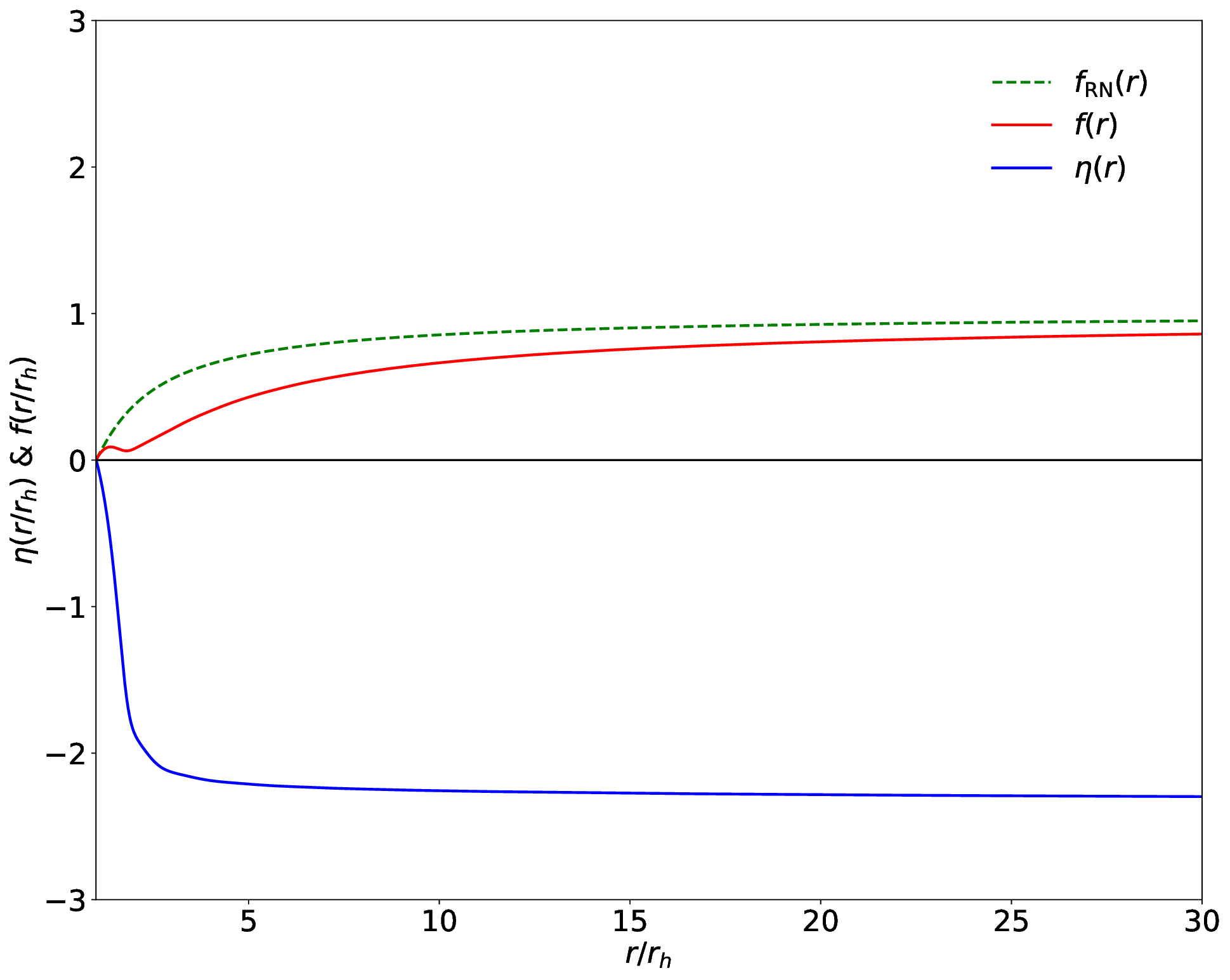}
 \caption{\footnotesize Radial profiles of metric potentials  $f(r)$ and $\eta(r)$ with $q=1.0$ and $\mu=0.1$ in units of $r_h$. The boundary conditions chosen are $V'(r_h)=1.0$ and $\psi(r_h)=0.5$. For comparison, the radial profile of the Reissner-Nordstr\"om metric potential $f_{\rm RN}(r)$ is also shown.}
 \label{fig.radprof_m_1}
\end{figure}

Figure \ref{fig.radprof_m_1} displays the radial profiles of the metric potentials $f(r)$ and $\eta_(r)$ with scalar charge $q=1.0$ and mass $\mu=0.1$, in units of $r_h$, with the horizon values taken as $V'(r_h)=1.0$, and $\psi_0=\psi(r_h)=0.5$. For comparison with the Reissner-Nordstr\"om solution, we have plotted the profile of $f_{\rm RN}(r)$ with the horizon condition $V'(r_h)=1.0$.

Our numerical data indicates that $f_{\rm RN}=0.95$ and $f=0.86$ at $\frac{r}{r_{\rm h}}=30$, $f_{\rm RN}=0.98$ and $f=0.94$ at $\frac{r}{r_{\rm h}}=100$, and  $f_{\rm RN}=0.99$ and $f=0.96$ at $\frac{r}{r_{\rm h}}=200$. Thus the metric  coefficient $f(r)$ approaches the limiting value of unity slowly (compared to the Reissner-Nordstr\"om case) due to backreaction of the charged scalar hair. It is however clear that the flat spacetime limit would be reached upon carrying out the numerical computation up to sufficiently  high values of the radial coordinate $r$.

As we can see from Figure \ref{fig.radprof_m_1}, the metric coefficient $f(r)$ is slightly non-monotonic near the horizon with a dip at about $r\sim2r_h$, and it acquires a profile similar to $f_{\rm RN}(r)$ for large $r$. We also see a steep fall in the metric potential $\eta(r)$ until about $r\sim2r_h$. This correlation originates from the coupled differential equations determining these metric coefficients. Afterwards, $\eta (r)$ appears to approach a constant negative value at infinity, indicating a re-parametrization of the time. These effects are solely due to the presence of the scalar hair.  In the absence of scalar hair, $\eta(r)$ would remain zero everywhere.

\begin{figure}[t!]
 \centering
 \includegraphics[width=.8\textwidth, height=9cm]{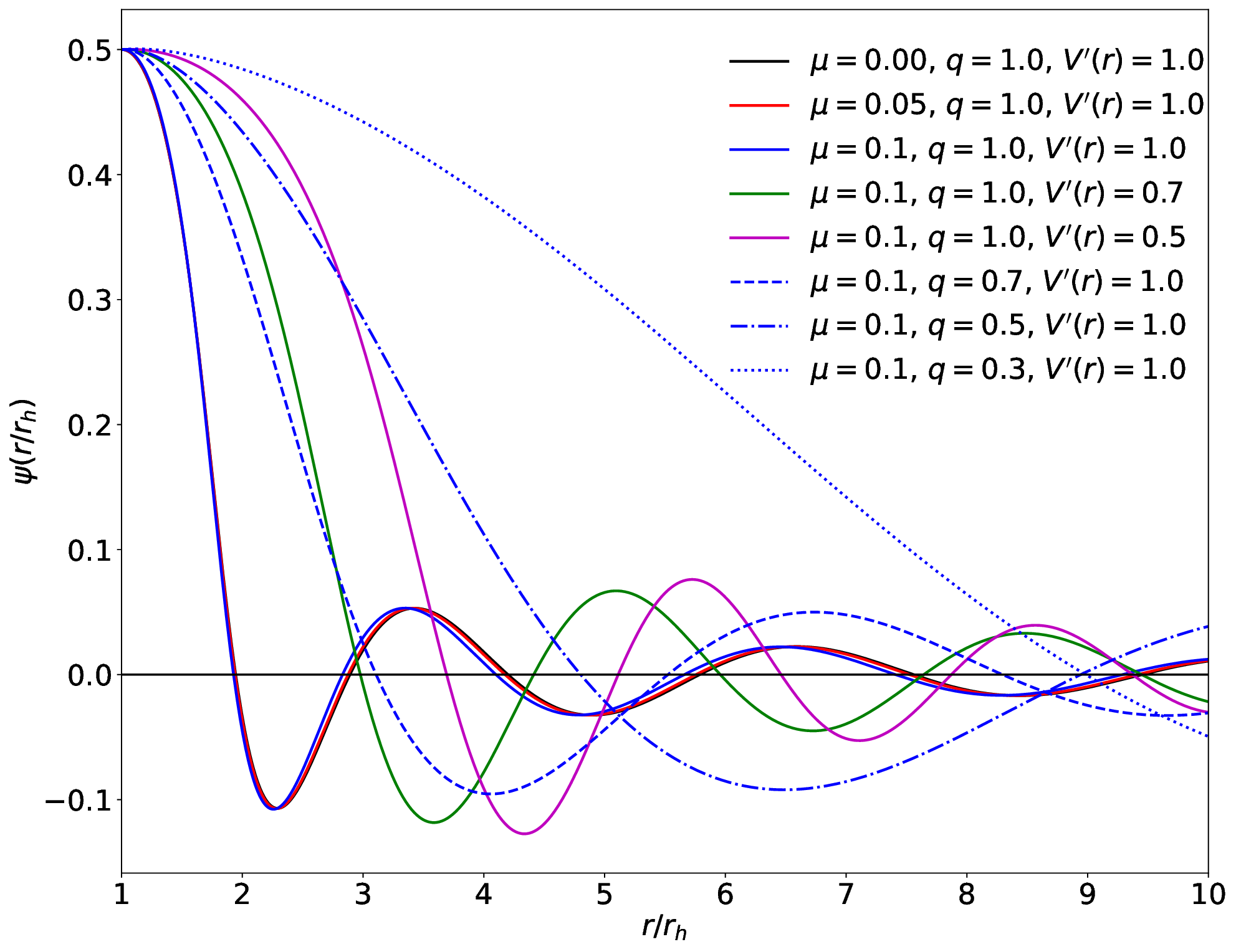}
 \caption{\footnotesize Radial profiles of hairy solutions $\psi(r)$ for different values of $V'(r_h)$, $q$ and $\mu$, with the horizon value $\psi(r_h)=0.5$.}
 \label{fig.psi_prof_fiff_dV}
\end{figure}

Figure \ref{fig.psi_prof_fiff_dV} shows the radial profiles of the scalar field $\psi(r)$ for different values of mass $\mu$ and charge $q$ of the scalar field, with $V'(r_h)=1.0$ and $\psi(r_h)=0.5$ at the horizon. All these profiles show a damped oscillatory behaviour with respect to the radial distance $r$. The profile with $\mu=0.1$ and $q=1.0$ shows a steep fall  until $r\sim2r_h$. We note that this steep fall in $\psi(r)$ is correlated with the non-monotonic behaviour of $f(r)$ and a steep fall of $\eta(r)$ as noted previously in Figure \ref{fig.radprof_m_1}.

It is clear from Figure \ref{fig.psi_prof_fiff_dV} that the hairy solution $\psi(r)$ approaches zero at infinity while performing a damped oscillation with increasing radial coordinate $r$. This creates a halo of charged cloud surrounding the black hole, the charge density approaching zero at infinity.

\begin{figure}[t!]
 \centering
 \includegraphics[width=.8\textwidth, height=9.25cm]{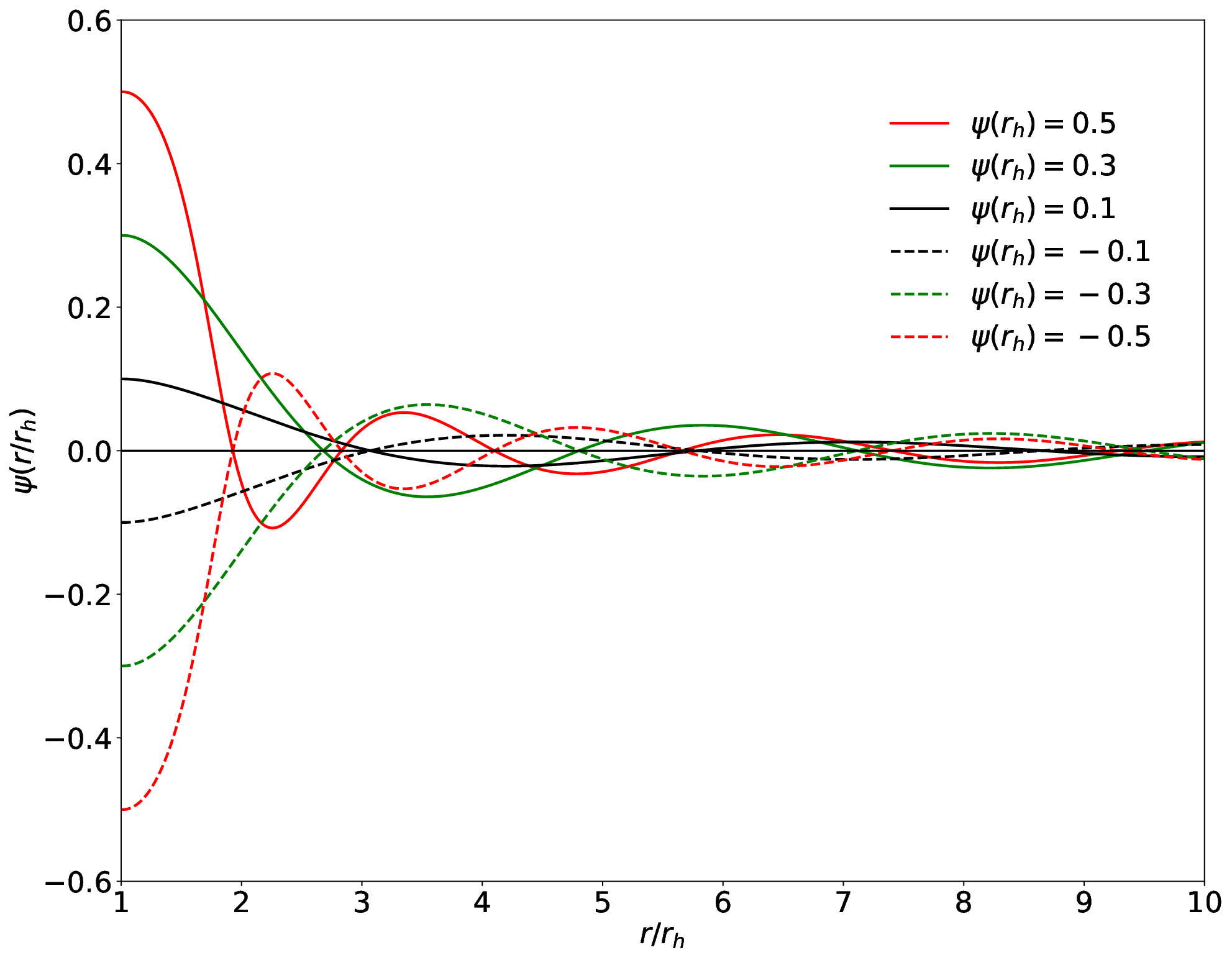}
 \caption{\footnotesize Radial profiles of $\psi(r)$ for different horizon values of $\psi(r_h)$, all having $V'(r_h)=1.0$, with scalar hair parameters $q=1.0$ and $\mu=0.1$. }
 \label{fig.phi_diff_IV}
\end{figure}

Figure \ref{fig.phi_diff_IV} shows radial profiles of the hairy solutions $\psi(r)$ with pair-wise positive and negative values of $\psi(r_h)$ at the horizon, with $V^{\prime}(r_h)=1.0$. Once again we can see damped oscillatory behaviours in all cases. An important observation here is that the solutions with pair-wise positive and negative values of $\psi(r_h)$ are exact mirror images of each other.

 \begin{figure}[b!]
 \centering
 \includegraphics[width=.8\textwidth, height=9.25cm]{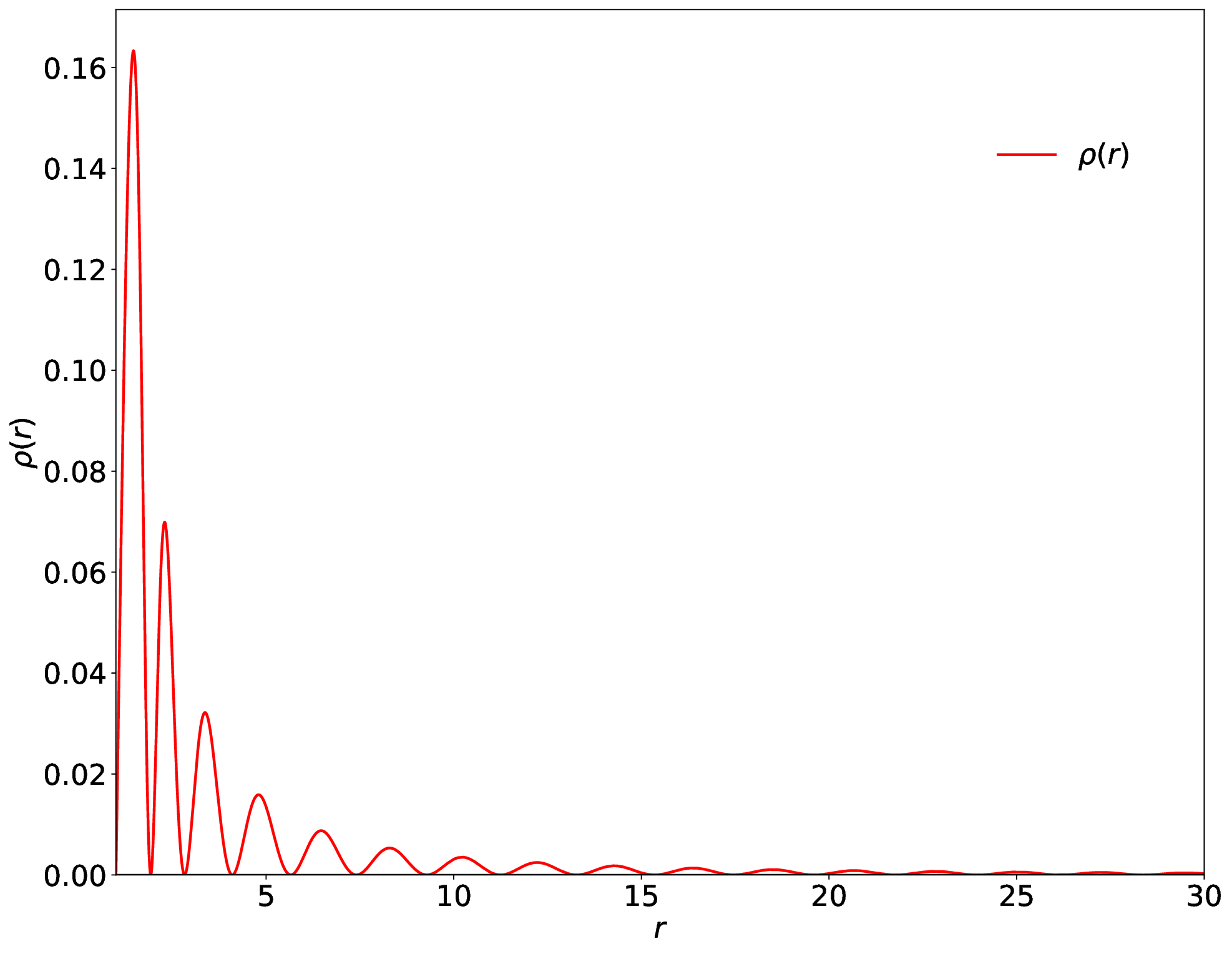}
 \caption{\footnotesize Radial profile of the charge density $\rho(r)=2q^{2}V\psi^{2}$ of the scalar hair with the choices $q=1.0$, $\mu=0.1$, $\psi(r_h)=0.5$ and $V'(r_h)=1.0$.}
 \label{fig.charge_den}
\end{figure}
The results shown in Figures \ref{fig.psi_prof_fiff_dV} and \ref{fig.phi_diff_IV} suggest possible existence of charged and massive scalar hair surrounding the charged black hole. Stability of such hairy solutions is discussed in Section \ref{sec_stability}.

It is also important to find the behaviour of the scalar charge density $ \rho(r)=2q^{2}V\psi^{2}$ surrounding the black hole. In Figure \ref{fig.charge_den}, we display the radial profile of $\rho(r)$ in one of the cases of our solutions. It is evident from the figure that the charge density $\rho(r)$ oscillates and falls off very rapidly with the radial coordinate $r$. We also note that there is a vanishing charge density at the horizon. The charge distribution appears to form concentric shells with varying charge densities that approach zero quickly with the shell radius. Practically, there would be no measurable charge density far from the black hole.

\begin{figure}[t!]
 \centering
 \includegraphics[width=.8\textwidth, height=9cm]{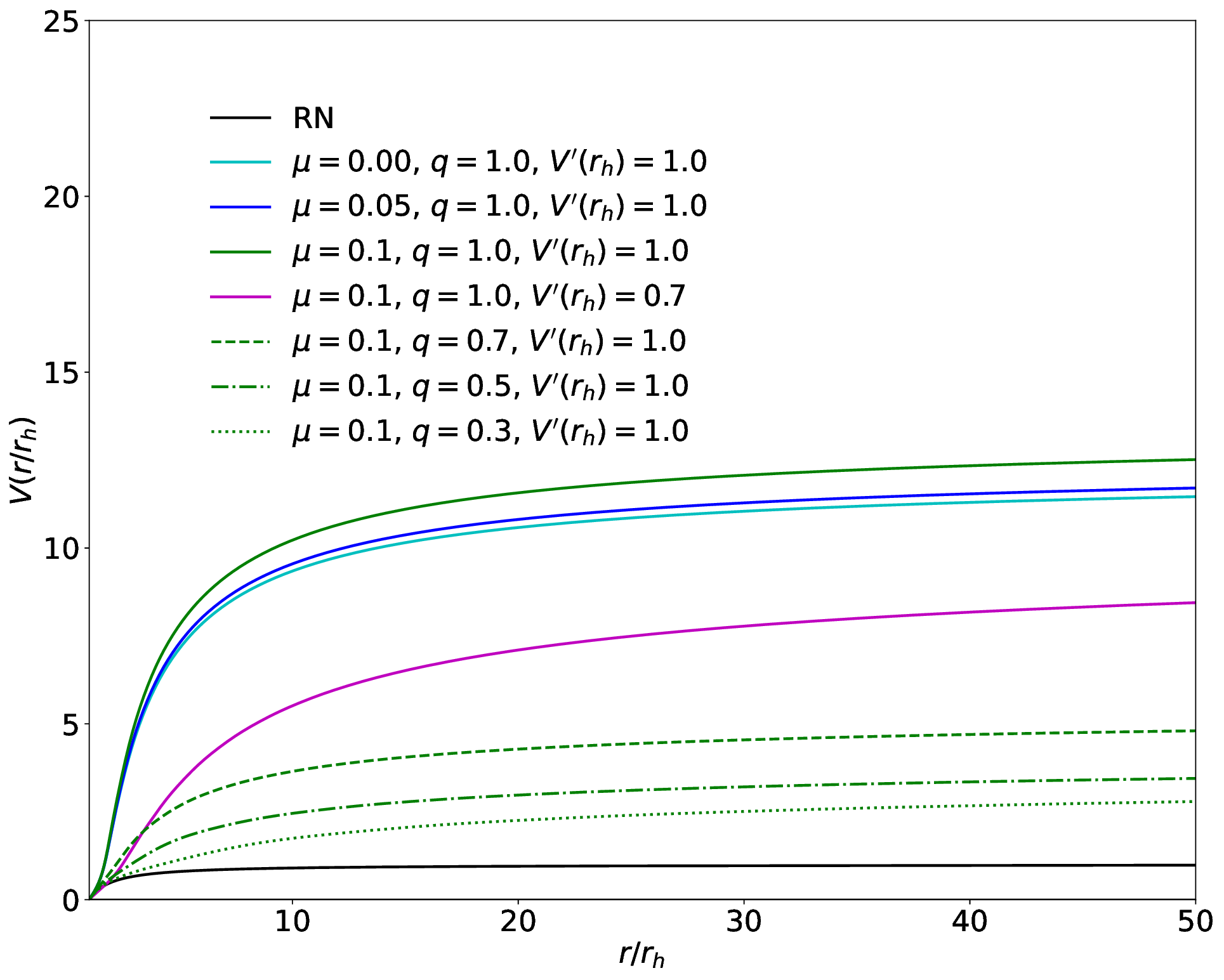}
 \caption{\footnotesize Radial profiles of the electrostatic potential $V(r)$ for different values of $V'(r_h)$ and scalar hair parameters $q$ and $\mu$, with the horizon condition $\psi(r_h)=0.5$. For comparison, the profile of the Reissner-Nordstr\"om potential is also shown. }
 \label{fig.v_comp_prof}
\end{figure}

 Figure \ref{fig.v_comp_prof} shows the radial profiles of the electrostatic potential $V(r)$ with different values of the parameters $q$ and $\mu$ of the scalar field, with horizon conditions $V'(r_h)=1.0$, and $\psi(r_h)=0.5$. All profiles of $V(r)$ show {\em monotonic} behaviours: the potential monotonically increases from zero at the horizon and appears to saturate to constant values at infinity. For the case $\mu$ fixed at $0.1$, as $q$ is increased, the profiles of $V(r)$ deviate further from the Reissner-Nordstr\"om profile; $q$ changing from $0.7$ to $1.0$ shows a rapid deviation, reaching a value much higher the Reissner-Nordstr\"om value at infinity. On the other hand, for the case $q$ fixed at $1.0$, the profiles undergo slight deviation as $\mu$ is increased from zero to $0.1$. It is obvious that the electrostatic potential $V(r)$ acquires additional contribution from the charged scalar hair so that it approaches a higher constant value at infinity. Moreover, the magnitude of the electric field $V^\prime(r)$ vanishes as $r\rightarrow\infty$, as expected.

 \section{Dynamical stability analyses}\label{sec_stability}
In order for the existence of the charged scalar hair surrounding a charged black hole, the hairy solution must be stable under time-dependant perturbation of the scalar field.

Consequently, we shall analyse the stability of the hairy solutions obtained in the previous sections. For definiteness, we shall carry out the stability analysis in two different ways. The first is based on Sturm-Liouville equation \cite{CS1, CS2,CS3, Bardeen_1966} whereas the second is based on a Schr\"odinger-like equation \cite{BRONNIKOV1978_1,Bronnikov1979_2} for the dynamic perturbation.

To carry out a dynamic stability analysis, we introduce time-dependence into the original scalar field, $\phi(r)\rightarrow\tilde{\phi}(r,t)$. The effect of this time-dependant field will make all the rest of the quantities time-dependant. We therefore make the changes: $f(r)\rightarrow\tilde{f}(r,t)$, $\eta(r)\rightarrow\tilde{\eta}(r,t)$ and $V(r)\rightarrow\tilde{V}(r,t)$. Accordingly, we have the modified metric as

\begin{equation}
 ds^{2}=-\tilde{f}(r,t)e^{-2\tilde{\eta}(r,t)}dt^{2}+\frac{dr^{2}}{\tilde{f}(r,t)}+r^{2}\left(d\theta^{2}+\sin^{2}\theta d\varphi^2\right).
\end{equation}

Employing the above metric in the field equations (\ref{eq_einstein_real}), (\ref{eq_KG_phi_1}), (\ref{eq_KG_phi_2}) and (\ref{eq_maxwell}), we obtain time-dependant differential equations. The Einstein field equation (\ref{eq_einstein_real}) leads to

\begin{equation}\label{eq_tilde_f_dash}
\begin{aligned}\tilde{f}' & =\frac{\left(1-\tilde{f}\right)}{r}-\frac{r}{2}\left[\frac{\left(\dot{\tilde{\psi}}^{2}+\dot{\tilde{\chi}}^{2}\right)}{fe^{-2\eta}}+\tilde{f}\left(\tilde{\psi}'^{2}+\tilde{\chi}'^{2}\right)+\frac{2q\tilde{V}}{\tilde{f}e^{-2\tilde{\eta}}}\left(\tilde{\psi}\dot{\tilde{\chi}}-\tilde{\chi}\dot{\tilde{\psi}}\right)\right.\\
 & \left.+e^{2\tilde{\eta}}\tilde{V}'^{2}+\left\{ \mu^{2}+\frac{q^{2}\tilde{V}^{2}}{\tilde{f}e^{-2\tilde{\eta}}}\right\} \left(\tilde{\psi}^{2}+\tilde{\chi}^{2}\right)\right],
\end{aligned}
\end{equation}

\begin{equation}\label{eq_tilde_h_dash}
\begin{aligned}\tilde{\eta}' & =-\frac{r}{2}\left[\frac{\left(\dot{\tilde{\psi}}^{2}+\dot{\tilde{\chi}}^{2}\right)}{\tilde{f}^{2}e^{-2\tilde{\eta}}}+\left(\tilde{\psi}'^{2}+\tilde{\chi}'^{2}\right)+\frac{q^{2}\tilde{V}^{2}}{\tilde{f}^{2}e^{-2\tilde{\eta}}}\left(\tilde{\psi}^{2}+\tilde{\chi}^{2}\right)+\frac{2q\tilde{V}}{\tilde{f}^{2}e^{-2\tilde{\eta}}}\left(\tilde{\psi}\dot{\tilde{\chi}}-\tilde{\chi}\dot{\tilde{\psi}}\right)\right],
\end{aligned}
\end{equation}
and

\begin{equation}\label{eq_tilde_f_dot}
\dot{\tilde{f}}=-r\tilde{f}\left[\tilde{\psi}'\dot{\tilde{\psi}}+\tilde{\chi}'\dot{\tilde{\chi}}+q\tilde{V}\left(\tilde{\psi}\tilde{\chi}'-\tilde{\chi}\tilde{\psi}'\right)\right].
\end{equation}

Similarly, the Maxwell equation (\ref{eq_maxwell}) yields

\begin{equation}\label{eq_tilde_maxwel_dash}
\begin{aligned}\tilde{V}'' & +\left(\tilde{\eta}'+\frac{2}{r}\right)\tilde{V}'-\frac{q^{2}\tilde{V}}{\tilde{f}}\left(\tilde{\psi}^{2}+\tilde{\chi}^{2}\right) -\frac{q}{\tilde{f}}\left(\tilde{\psi}\dot{\tilde{\chi}}-\tilde{\chi}\dot{\tilde{\psi}}\right)=0
\end{aligned}
\end{equation}
and

\begin{equation}
 \dot{\tilde{V}}'+\tilde{V}'\dot{\tilde{\eta}}-q\tilde{f}e^{-2\tilde{\eta}}\left(\tilde{\psi}\tilde{\chi}'-\tilde{\chi}\tilde{\psi}'\right)=0.
\end{equation}

In addition, the Klein-Gorden equations (\ref{eq_KG_phi_1}) and (\ref{eq_KG_phi_2}) are now modified to

\begin{equation}\label{eq_tilde_kg_dash1}
\begin{aligned}\ddot{\tilde{\psi}}-\tilde{f}^{2}e^{-2\tilde{\eta}}\tilde{\psi}'' & -\dot{\tilde{\psi}}\left(\frac{\dot{\tilde{f}}}{\tilde{f}}-\dot{\tilde{\eta}}\right)-\left\{ \frac{\tilde{f}'}{\tilde{f}}-\tilde{\eta}'+\frac{2}{r}\right\} \tilde{f}^{2}e^{-2\tilde{\eta}}\tilde{\psi}'-q\tilde{\chi}\dot{\tilde{V}}\\
 & +q\tilde{\chi}\tilde{V}\left(\frac{\dot{\tilde{f}}}{\tilde{f}}-\dot{\tilde{\eta}}\right)-2q\tilde{V}\dot{\tilde{\chi}}-q^{2}\tilde{V}^{2}\tilde{\psi}+\mu^{2}\tilde{f}e^{-2\tilde{\eta}}\tilde{\psi}=0
\end{aligned}
\end{equation}
and

\begin{equation}\label{eq_tilde_kg_dash2}
\begin{aligned}\ddot{\tilde{\chi}}-\tilde{f}^{2}e^{-2\tilde{\eta}}\tilde{\chi}'' & -\dot{\tilde{\chi}}\left(\frac{\dot{\tilde{f}}}{\tilde{f}}-\dot{\tilde{\eta}}\right)-\left\{ \frac{\tilde{f}'}{\tilde{f}}-\tilde{\eta}'+\frac{2}{r}\right\} \tilde{f}^{2}e^{-2\tilde{\eta}}\tilde{\chi}'+q\tilde{\psi}\dot{\tilde{V}}\\
 & -q\tilde{\psi}\tilde{V}\left(\frac{\dot{\tilde{f}}}{\tilde{f}}-\dot{\tilde{\eta}}\right)+2q\tilde{V}\dot{\tilde{\psi}}-q^{2}\tilde{V}^{2}\tilde{\chi}+\mu^{2}\tilde{f}e^{-2\tilde{\eta}}\tilde{\psi}=0.
\end{aligned}
\end{equation}

In order to carry out the stability analysis, we assume that the time-dependant perturbations are sinusoidal, and we perturb the time-dependant fields around the static solutions as
\begin{equation}\label{eqinusoidal}
\begin{aligned}\tilde{f}(r,t) & =f(r)+\varepsilon f_{1}(r)e^{-i\Omega t},\\
\tilde{\eta}(r,t) & =\eta(r)+\varepsilon \eta_{1}(r)e^{-i\Omega t},\\
\tilde{V}(r,t) & =V(r)+\varepsilon V_{1}(r)e^{-i\Omega t},\\
\tilde{\psi}(r,t) & =\psi(r)+\varepsilon\psi_{1}(r)e^{-i\Omega t},\\
\tilde{\chi}(r,t) & =\chi(r)+\varepsilon\chi_{1}(r)e^{-i\Omega t},
\end{aligned}
\end{equation}
where $\varepsilon$ is a very small quantity ($\varepsilon\ll1$). In this Section, quantities carrying a subscript 1 indicates the perturbation field depending on the radial coordinate, and $\Omega$ is the eigen frequency to be determined from the stability analysis.

Substituting the above expressions (\ref{eqinusoidal}) in (\ref{eq_tilde_f_dash}), (\ref{eq_tilde_h_dash}), (\ref{eq_tilde_f_dot}), (\ref{eq_tilde_maxwel_dash}) and (\ref{eq_tilde_kg_dash1}), we obtain the following set of equations:
\begin{equation}
 f_{1}=-2fr\psi'\psi_{1},
 \end{equation}

\begin{equation}
\eta_{1}'=2r\left[\frac{q^{2}V^{2}}{f^{2}e^{-2\eta}}\psi\psi_{1}-\psi'\psi_{1}'\right],
\end{equation}

\begin{equation}
 V_{1}'=-V'\eta_{1},
\end{equation}
and
\begin{equation}\label{eq_eigen}
\begin{aligned}f^{2}e^{-2\eta}r\psi_{1}''+ & f^{2}e^{-2\eta}\left\{ \frac{f'}{f}-\eta'+\frac{2}{r}\right\} r\psi_{1}'+f^{2}e^{-2\eta}\left(\frac{f'}{f}-\eta'\right)\psi_{1}\\
-\left\{ \frac{f^{2}e^{-2\eta}}{r}\right. & \left(\frac{f'}{f}-\eta'\right)+\mu^{2}fe^{-2\eta}+3q^{2}V^{2}-2rfe^{-2\eta}f'\psi'^{2}-2f^{2}e^{-2\eta}\psi'^{2}\\
+4\mu^{2}rfe^{-2\eta} & \psi'\psi-2r^{2}f^{2}e^{-2\eta}\psi'^{4}\left.-2r^{2}q^{2}V^{2}\psi^{2}\psi'^{2}\right\} r\psi_{1}=-\Omega^{2}r\psi_{1}.
\end{aligned}
\end{equation}

To obtain the scalar perturbation equation in the Sturm-Liouville form, we define $\xi=r\psi_{1}$, leading to
\begin{equation}\label{eq_SL}
\begin{aligned}\left(fe^{-\eta}\xi'\right)' & -\left\{ \frac{1}{r}\left(fe^{-\eta}\right)^{'}+\mu^{2}e^{-\eta}+\frac{3q^{2}V^{2}}{fe^{-\eta}}-2re^{-\eta}f'\psi'^{2}-2fe^{-\eta}\psi'^{2}\right.\\
 & +4\mu^{2}re^{-\eta}\psi'\psi-2r^{2}fe^{-\eta}\psi'^{4}\left.-2r^{2}\frac{q^{2}V^{2}}{fe^{-\eta}}\psi^{2}\psi'^{2}\right\} \xi=-\frac{\Omega^{2}}{fe^{-\eta}}\xi.
\end{aligned}
\end{equation}

Equation (\ref{eq_SL}) is in the form of an eigenvalue equation. We notice from (\ref{eqinusoidal}) that the exponential terms will have a growing mode if $\Omega^2<0$. Writing $\Omega=\pm i\alpha$, with $\alpha>0$, the eigen solutions will have an exponentially growing mode $e^{\alpha t}$ and a decaying mode $e^{-\alpha t}$. The exponentially growing mode $e^{\alpha t}$ makes the solution unstable. On the other hand, for the case $\Omega^2>0$, we may write $\Omega=\pm \omega$, so that both modes $e^{i\omega t}$ and $e^{-i\omega t}$ perform stable sinusoidal oscillations.

Thus the sign of the eigenvalue $\Omega^2$ dictates the stability of the hairy solution. Solutions with $\Omega^2>0$ are stable, whereas $\Omega^2<0$ corresponds to instability, and $\Omega^2=0$ signifies onset of instability.

\begin{figure}
 \centering
 \includegraphics[width=.8\textwidth, height=9cm]{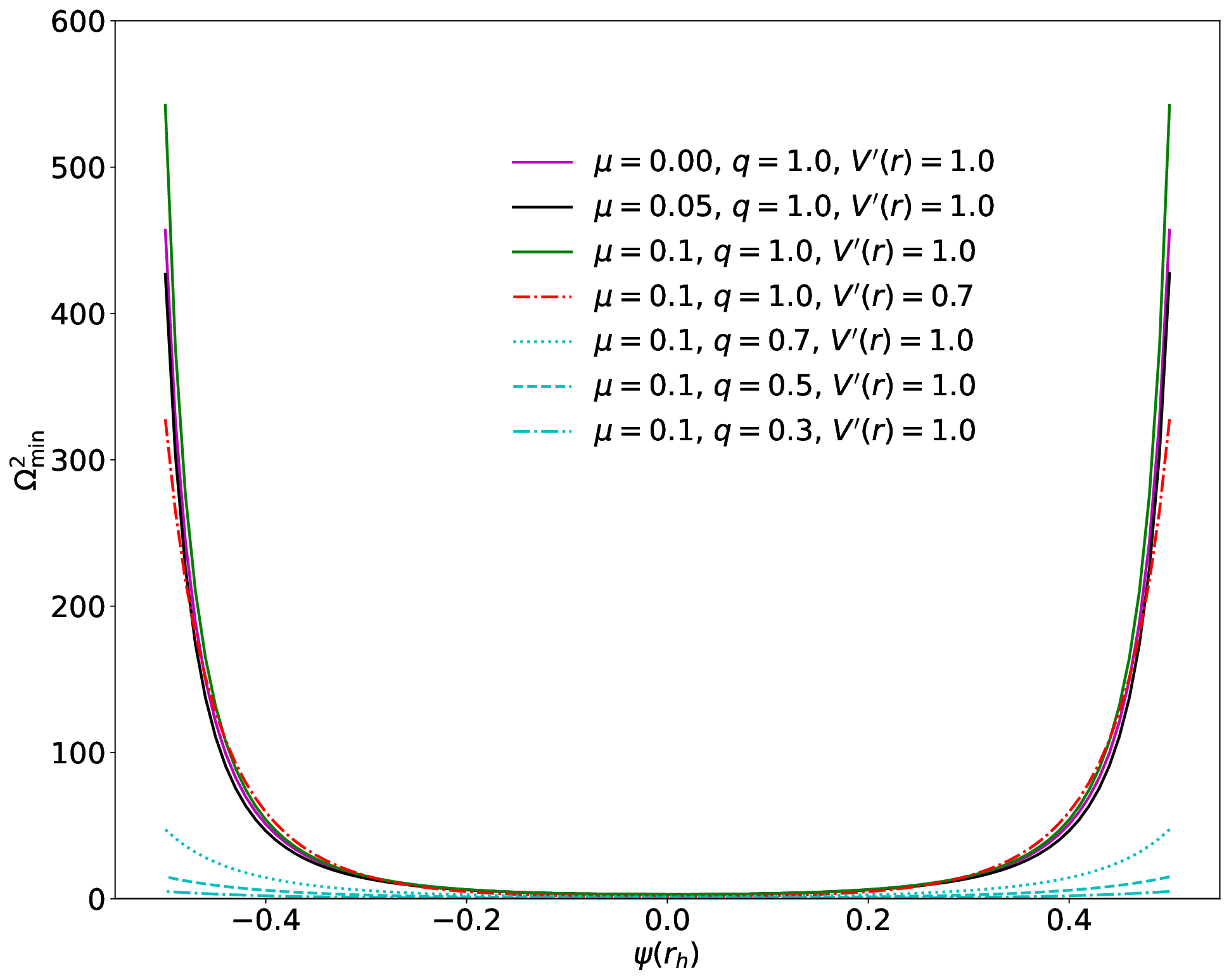}
 \caption{\footnotesize The minimum Sturm-Liouville eigenvalue $\Omega_{\rm min}^2$ with respect to horizon value $\psi(r_h)$ with the choice of trial function $\xi(x)=a\tanh(x)$.}
 \label{fig.Omega_Ph}
\end{figure}

\subsection{Analysis of Sturm-Liouville equation}

As noted earlier, Equation (\ref{eq_SL}) can be expressed in the Sturm-Liouville form,
\begin{equation}\label{eq_SL_real}
 \frac{d}{dr}\left(P(r)\frac{d\xi(r)}{dr}\right)+Q(r)\xi(r)+\Omega^2R(r)\xi(r)=0,
\end{equation}
where
\begin{equation}
\begin{aligned}P(r) & =fe^{-\eta},\\
Q(r) & =-\left[\frac{1}{r}\left(fe^{-\eta}\right)^{'}+\mu^{2}e^{-\eta}+\frac{3q^{2}V^{2}}{fe^{-\eta}}-2re^{-\eta}f'\psi'^{2}-2fe^{-\eta}\psi'^{2}\right.\\
 & \left.+4\mu^{2}re^{-\eta}\psi'\psi-2r^{2}\left(fe^{-\eta}\psi'^{4}+\frac{q^{2}V^{2}}{fe^{-\eta}}\psi^{2}\psi'^{2}\right)\right],\\
R(r) & =\frac{1}{fe^{-\eta}}.
\end{aligned}
\end{equation}

The smallest eigenvalue $\Omega_{\rm min}^2$ of the Sturm-Liouville equation (\ref{eq_SL_real}) can be obtained by employing the variational principle, giving

\begin{equation}\label{eq_ev_min}
\Omega_{{\rm min}}^{2}=\stackrel{\rm min}{\mbox{}_{\xi(r)}}\frac{\int_{r_{h}}^{\infty}dr\ P(r)\xi'^{2}(r)-\int_{r_{h}}^{\infty}dr\ Q(r)\xi^{2}(r)}{\int_{r_{h}}^{\infty}dr\ R(r)\xi^{2}(r)},          \end{equation}
which can be evaluated with a trial function for $\xi(r)$.

For definiteness, we choose two different trial functions, $\xi(x)=a\tanh x$ and $\xi(x)=bxe^{-x}$, where $x=\frac{r - r_h}{r_h}$ and $a$ and $b$ are disposable constants. With these trial functions, we carry out the integrations numerically in Equation (\ref{eq_ev_min}) to obtain the minimum eigenvalue $\Omega_{\rm min}^{2}$.

Figure \ref{fig.Omega_Ph} shows the behaviours of $\Omega_{\rm min}^{2}$ with respect to $\psi(r_h)$, with different values of the parameters $q$ and $\mu$, employing the trial function $\xi=a\tanh x$. It is clear from the figure that $\Omega_{\rm min}^{2}$ is positive and non-zero over a continuous range of horizon conditions $\psi(r_h)$ involving both positive and negative values.

\begin{figure}[t!]
 \centering
 \includegraphics[width=.8\textwidth, height=9cm]{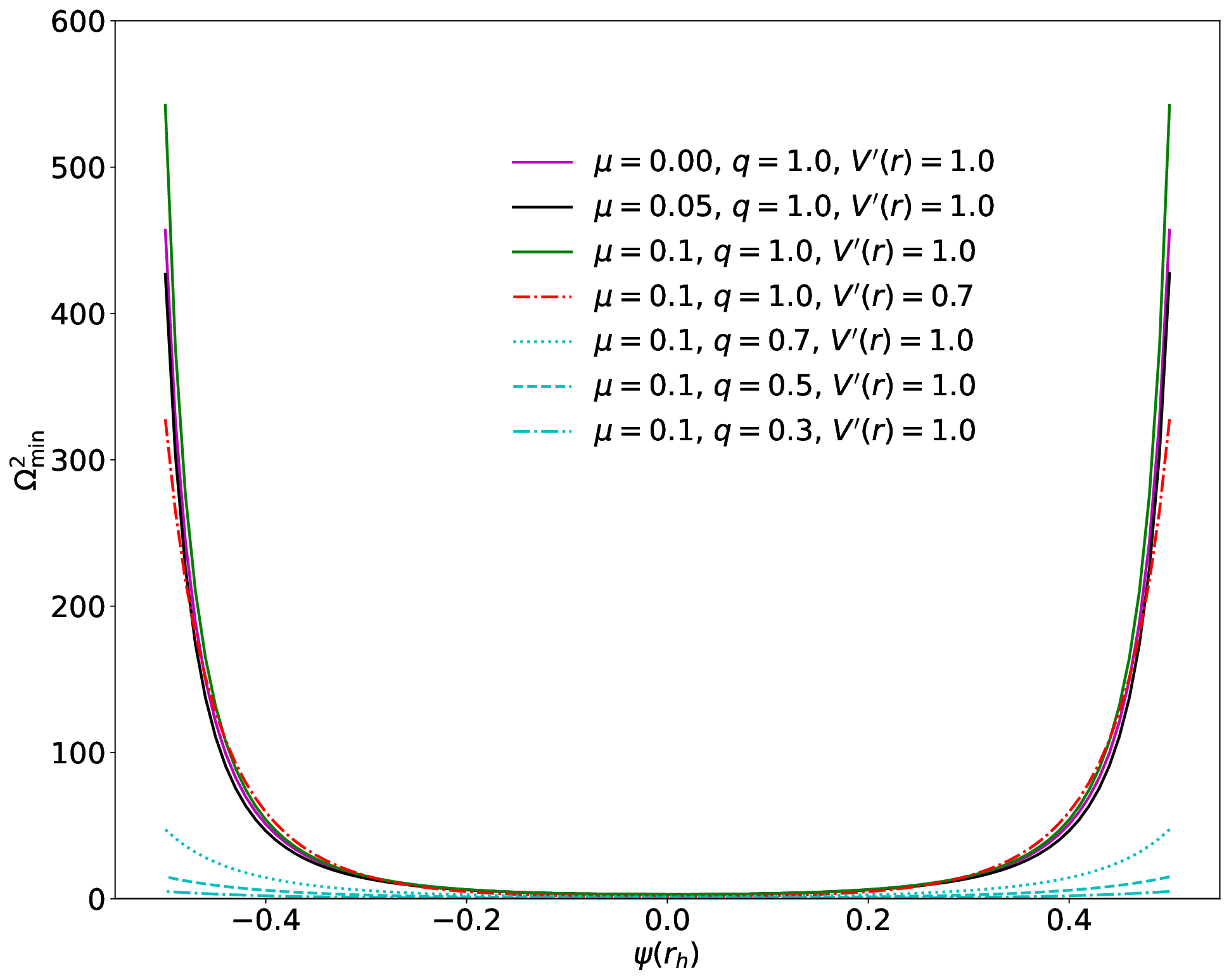}
 \caption{\footnotesize The minimum Sturm-Liouville eigenvalue $\Omega_{\rm min}^2$ with respect to horizon value $\psi(r_h)$ with the choice of trial function $\xi(x)=bxe^{-x}$.}
 \label{fig.2Omega_Ph}
\end{figure}

We confirm the above results with another choice of trial function for $\xi(r)$. Figure \ref{fig.2Omega_Ph} shows the behaviours of $\Omega_{\rm min}^{2}$ with respect to $\psi(r_h)$ for different values of the parameters $q$ and $\mu$, with the choice of trial function $\xi=bxe^{-x}$. We once again see that $\Omega_{\rm min}^{2}$ turns out to be positive and non-zero over a continuous range of positive and negative horizon conditions for $\psi(r_h)$.

In order to analyse whether $\Omega_{\rm min}^{2}$ can become negative, we see from Figures \ref{fig.Omega_Ph} and \ref{fig.2Omega_Ph} that such possibility may exist for the initial condition $\psi(r_h)=0$. We therefore have to analyse equation (\ref{eq_SL}) at the horizon, $r=r_h$. Equation (\ref{eq_near_hor}) indicates that $\psi^\prime(r_h )$ is proportional to $\psi(r_h)$. The analysis in the paragraph after equation (\ref{eq_EM_tensor_mixed}) shows that$ V^2(r_h)/f(r_h) =0$. Thus, at the horizon equation (\ref{eq_SL}) reduces to $\left(fe^{-\eta}\xi'\right)'-\left\{ \frac{1}{r}\left(fe^{-\eta}\right)^{\prime}+\mu^{2}e^{-\eta}\right\} \xi=-\frac{\Omega^{2}}{fe^{-\eta}}\xi$. Since $\eta(r_h) =0$, this equation further reduces to  $f^{2}\eta^{\prime}\xi^{\prime}-f^{2}\xi^{\prime\prime}+f\left\{ \frac{1}{r}f^{^{\prime}}-\frac{1}{r}f\eta^{\prime}+\mu^{2}\right\} =\Omega^{2}$. Moreover, with the boundary condition $\xi(r_h)=0$ as well as $\xi'(r_h)=0$, this equation further reduces to $f\left\{ -f\xi^{\prime\prime}+\frac{1}{r}f^{^{\prime}}-\frac{1}{r}f\eta^{\prime}+\mu^{2}\right\} =\Omega^{2}$. This implies $\Omega^2_{\rm min}=0$ since $f(r_h)=0$. Thus, $\Omega_{\rm min}^{2}$  cannot become negative for the initial condition $\psi(r_h)=0$. For other initial conditions with $\psi(r_h)\neq0$ we can infer from Figures \ref{fig.Omega_Ph} and \ref{fig.2Omega_Ph} that  $\Omega_{\rm min}^{2}>0$.

The above results indicate that $\Omega_{\rm min}^{2}>0$ over a continuous range of horizon values $\psi(r_h)$, both positive and negative. It may therefore be concluded that the scalar hair solutions remain stable with respect to sinusoidal perturbations.

We further confirm this conclusion employing another method for stability analysis in the next Subsection.

\begin{figure}[t!]
 \centering
 \includegraphics[width=.8\textwidth, height=9cm]{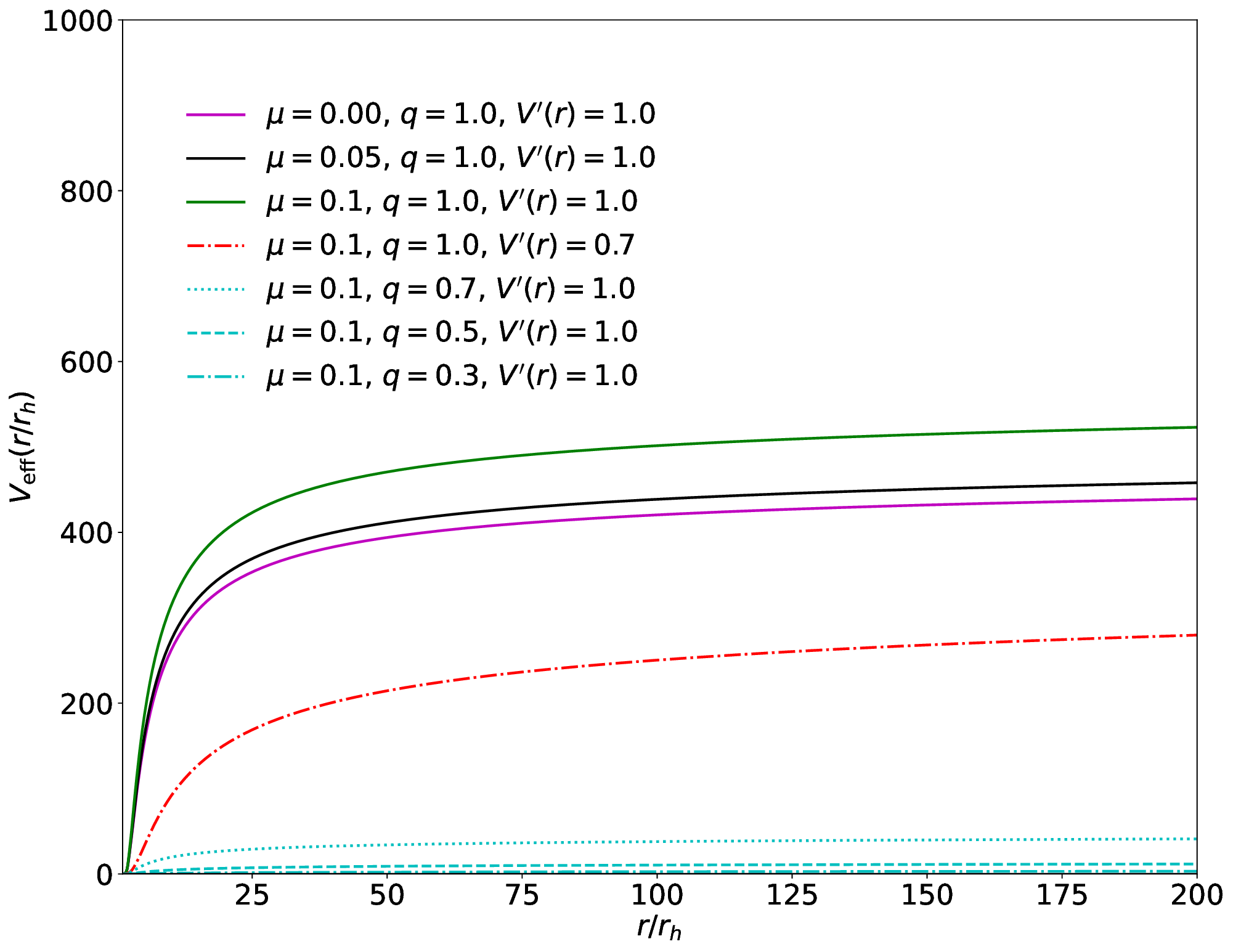}
 \caption{\footnotesize Radial profiles of the effective potential $V_{\rm eff}(r)$ with horizon values $\psi(r_h)=0.5$ and $V'(r_h)=1.0$, making different choices for the scalar hair parameters $q$ and $\mu$.}
 \label{fig.v_eff_plot}
\end{figure}

\subsection{Analysis of Schr\"odinger-like equation}

Transforming to the tortoise coordinate $r^*$, defined by $dr^* = \frac{e^{\eta}}{f}dr$, Equation (\ref{eq_SL}) can be reduced to a Schr\"odinger-like equation,

\begin{equation}\label{eq_schrodinger_like}
 -\frac{d^{2}\xi}{dr^{*2}}+V_{\rm eff}\xi=\Omega^{2}\xi,
\end{equation}
where the effective potential $V_{{\rm eff}}$ is defined as
\begin{equation}\label{eq_veff}
\begin{aligned}V_{{\rm eff}} & =\frac{fe^{-\eta}}{r}\left(fe^{-\eta}\right)^{'}+\mu^{2}fe^{-2\eta}+3q^{2}V^{2}-2rfe^{-2\eta}f'\psi'^{2}-2f^{2}e^{-2\eta}\psi'^{2}\\
 & \ +4\mu^{2}rfe^{-2\eta}\psi'\psi-2r^{2}f^{2}e^{-2\eta}\psi'^{4}-2r^{2}q^{2}V^{2}\psi^{2}\psi'^{2}.\
\end{aligned}
\end{equation}

 Left-multiplying Equation (\ref{eq_schrodinger_like}) with the conjugate $\bar{\xi}$ and integrating from the horizon ($-\infty$ in the tortoise coordinate) to infinity, we obtain

\begin{equation}\label{eq_schrodinger_like_int_1}
\begin{aligned}-\left[\bar{\xi}\frac{d\xi}{dr^{*}}\right]_{-\infty}^{\infty} & +\int_{-\infty}^{\infty}dr^{*}\left[\left|\frac{d\xi}{dr^{*}}\right|^{2}+V_{\rm eff}\left|\xi\right|^{2}\right]\\
 & =\Omega^{2}\int_{-\infty}^{\infty}dr^{*}\left|\xi\right|^{2}.
\end{aligned}
\end{equation}

The first term on the left hand side vanishes upon imposing the boundary conditions $\bar{\xi}\rightarrow0$ as $r^*\rightarrow\pm\infty$. This leads to
\begin{equation}\label{eq_schrodinger_like_int_2}
\int_{-\infty}^{\infty}dr^{*}\left[\left|\frac{d\xi}{dr^{*}}\right|^{2}+V_{\rm eff}\left|\xi\right|^{2}\right]=\Omega^{2}\int_{-\infty}^{\infty}dr^{*}\left|\xi\right|^{2}.
\end{equation}

It is evident from Equation (\ref{eq_schrodinger_like_int_2}) that if the profile of $V_{\rm eff}$ stays always positive, we must have $\Omega^2>0$. This follows from the fact that all remaining quantities in (\ref{eq_schrodinger_like_int_2}) are positive. On the other hand, if $V_{\rm eff}$ is negative in some regions, $\Omega^2$ can either be positive or negative depending upon whether the contributions from the negative regions are sufficiently small or sufficiently large, respectively.

\begin{figure}[t!]
 \centering
 \includegraphics[width=.8\textwidth, height=9cm]{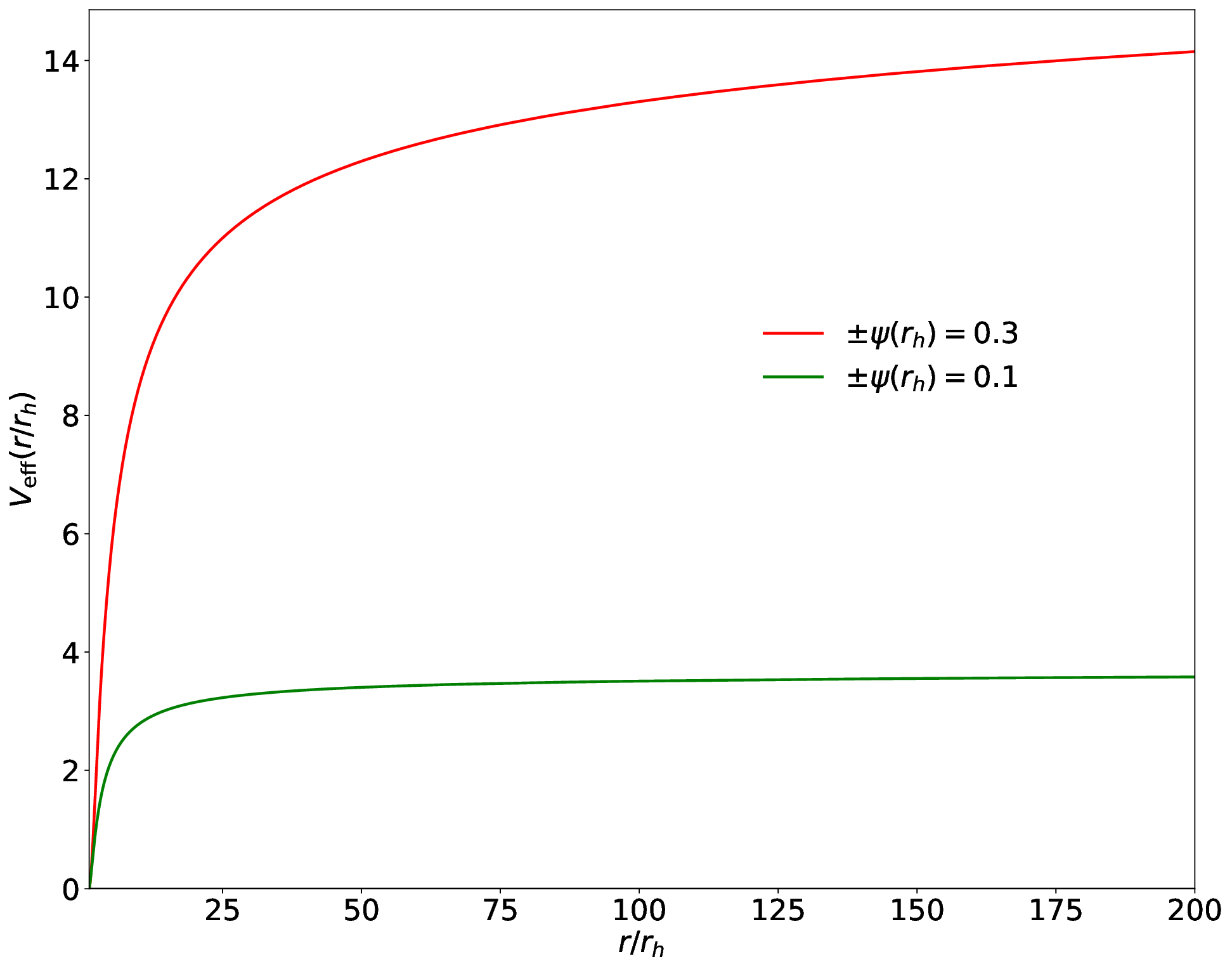}
 \caption{\footnotesize Radial profiles of the effective potential $V_{\rm eff}(r)$ with pair-wise positive and negative horizon values of $\psi(r_h)$, all having $V'(r_h)=1.0$, with scalar hair parameters, $q=1.0$ and $\mu=0.1$. }
 \label{fig.v_eff_plot_ph}
\end{figure}

We therefore plot $V_{\rm eff}(r)$ with different values for the parameters $q$ and $\mu$. Figure \ref{fig.v_eff_plot} shows the profiles of $V_{\rm eff}(r)$ with the horizon condition $\psi(r_h)=0.5$. Moreover, Figure \ref{fig.v_eff_plot_ph} shows the profiles of $V_{\rm eff}(r)$ for pair-wise positive and negative values of $\psi(r_h)$, each with $q=1.0$ and $\mu=0.1$. Interestingly, the curves for every pair overlap exactly, implying that the stability of the solution is unaffected by flipping the phase of $\psi(r)$ at the horizon.

Remarkably, all these profiles of $V_{\rm eff}(r)$ are positive everywhere, and therefore they imply that $\Omega^2>0$, so that the corresponding hairy solutions are stable. This is consistent with our previous stability analysis based on the Sturm-Liouville equation.

\section{Discussion and conclusion}\label{sec_conc}
As we have noted, the no-hair theorem forbids any property other than mass, charge and angular momentum to parametrize a black hole. In this regard, investigations on the possibility of existence of a scalar hair surrounding the black hole was first considered  by Bekenstein \cite{BEKENSTEIN1974,BEKENSTEIN1975} with a scalar field conformally coupled to the geometry of a charged black hole. In addition, as discussed earlier, Mayo and Bekenstein \cite{Mayo_Bekenstein_1996} considered the possibility of having hairy solutions with both electrically neutral and charged scalar fields having conformal coupling with the gravitational field of a charged black hole.

Subsequently, although charged hairy solutions were obtained in Mayo and Bekenstein's model with polynomial scalar potential of sixth degree and the scalar charge much smaller than the black hole charge, a natural question arises as to whether a charged scalar hair would {\em exist} if the scalar field is taken to be a charged massive Klein-Gordon field (with quadratic potential) without any limitation on  its charge.

We note that the {\em existence} of a hairy solution can be established if  the following two conditions are satisfied: (i) the necessary condition of meeting all relevant energy conditions supplemented with the causality condition, and (ii) the sufficient condition that the hairy solution is {\em stable} against time dependant perturbations.

In this context, we considered in this paper the Einstein-Maxwell-charged scalar field model with a spherically symmetric charged scalar field electromagnetically coupled to a charged black hole. The electromagnetic coupling naturally follows from gauge coupling (via covariant derivative) of the charged scalar field. We also considered both massless and massive cases of the charged scalar field.

In this Einstein-Maxwell-charged scalar framework, we found that the Einstein field equations, Maxwell's equations, and the Klein-Gorden equation are substantially modified by acquiring additional terms coming from the charged scalar field. These equations form a coupled set of nonlinear differential equations.

Before proceeding any further, it was important to ascertain whether the necessary condition (i) of all relevant energy conditions supplemented by the causality condition are obeyed by the Einstein-Maxwell-charged scalar field model. We therefore considered all energy conditions following from the mixed components of the energy-momentum tensor and the causality condition. Upon expressing the mixed components of the energy-momentum tensor in terms of the scalar field and other quantities, we found that all energy conditions coupled with the causality condition are satisfied. This suggests the possibility of existence of a charged hairy solution.

Thus our next task was to find charged hairy solutions from the field equations. Since the field equations consist of a set of {\it nonlinear} coupled differential equations, any analytical attempt to solve them was not possible. Consequently, we solved the field equations by exact numerical integration, with appropriate boundary conditions at the horizon, and with different values of the electric charge $q$ and mass $\mu$ of the scalar field.

Our numerical integration showed oscillatory solutions for the scalar field which damps out with the radial distance. We also studied the behaviour of the charge density $\rho(r)$ of the scalar hair. We found that its radial profile makes a series of bounces with heights falling off rapidly with the radial coordinate. This structure indicates that the scalar charge is distributed in concentric shells around the black hole. The rapid falloff of the charge density indicates that there would be practically no measurable charge at a large distance from the black hole.

Our solutions further indicate that the electric potential $V(r)$ surrounding the black hole is a monotonic function of the radial distance $r$. This behaviour is consistent with the requirement from the Maxwell's equation.

Furthermore, the electric potential can take highly enhanced values compared to the normal Reissner-Nordstr\"om case without a hair. This is due to additional contribution to the electric potential coming from the charged scalar hair in addition to that from the charged black hole.

Although we obtained the charged hairy solutions by exact numerical integration, their existence can be confirmed only by stability analysis with respect to time dependant perturbations about the static hairy solutions. Consequently, we carried out dynamical stability analyses by perturbing the static solution with a sinusoidal perturbation. For definiteness in the result, we employed two different methodologies of stability analyses.

In the first methodology, we obtained the minimum frequency of the eigenmodes of the Sturm-Liouville equation for the dynamics of the radial perturbation. We found that the minimum eigen frequency turns out to be real-valued for our solutions, indicating their stability.

In the second methodology, we confirmed the above stability analysis. We converted the dynamical equation of the perturbation into a Schr\"odinger-like equation with an effective potential determined by the static hairy solution. An analysis of this Schr\"odinger equation indicated that the eigen frequency will be real-valued, giving stability of the solution, provided the effective potential remains positive everywhere. Upon plotting the effective potential using our different static solutions, we found that the effective potential remains positive everywhere, indicating real valuedness of all eigen frequencies for the static solutions. This confirms our previous stability analysis based on the determination of minimum eigen frequency of the Sturm-Liouville equation.

We note in passing that in references \cite{Hong_suzuki_yamada_PLB, Hong_suzuki_yamada_PRL, Herdeiro_radu_2020} the potential of the scalar field was considered to be a polynomial of degree 6, with only even powers of $\phi$, having the form $V(\phi)=a \phi^2 - b\phi^4 + c\phi^6 $. On the other hand, the potential in the present work is only quadratic, that is $V(\phi)=a\phi^2$. This difference in the forms of the potentials makes a crucial change in the solutions. In references \cite{Hong_suzuki_yamada_PLB, Hong_suzuki_yamada_PRL, Herdeiro_radu_2020}, a non-oscillatory decaying solution was obtained in the polynomial model for the potential with specific choices for the parameters. On the other hand, we obtained a decaying oscillatory solution for the scalar hair with the quadratic potential. In addition, we confirmed the stability of our decaying oscillatory solutions.

The fact that all energy conditions coupled with the causality condition are satisfied, in addition to satisfying the condition of stability of the static hairy solutions with respect to dynamical perturbations, suggest that a Reissner-Nordstr\"om black hole can be surrounded by a stable and electrically charged scalar hair with quadratic scalar potential.

In this context, we note that our theory considers {\em static} hairy solutions and confirms their stability while it does not include the dynamics of how such hairy solutions form around the black hole. It is well-known that a Reissner-Nordstr\"om black hole may undergo the phenomenon of  superradiance instability due to scattering of incident charged particles, giving rise to outgoing charges if the condition of superradiance is fulfilled \cite{Bekenstein1973}. However, for an asymptotically flat Reissner-Nordstr\"om spacetime, the necessary conditions required to trigger a possible superradiance instability cannot be satisfied \cite{HOD3,HOD1}. Thus the origin of charged scalar hair in our model does not lie in the phenomenon of superradiance instability and it remains a matter of exploration in the future.

\section*{Acknowledgments}
Muhammed Shafeeque is indebted to the Ministry of Education, Government of India, for financial assistance through a Research Fellowship. The Authors would like to thank the Indian Institute of Technology Guwahati for providing access to computing and supercomputing facilities.

\providecommand{\href}[2]{#2}\begingroup\raggedright\endgroup


\begin{thebibliography}{10}

\bibitem{NH1_israel67}
W.~Israel, \emph{Event horizons in static vacuum space-times},
  \href{https://doi.org/10.1103/PhysRev.164.1776}{\emph{Phys. Rev.} {\bfseries
  164} (1967) 1776}.

\bibitem{NH2_Israel1968}
W.~Israel, \emph{Event horizons in static electrovac space-times},
  \href{https://doi.org/10.1007/BF01645859}{\emph{Communications in
  Mathematical Physics} {\bfseries 8} (1968) 245}.

\bibitem{NH3_carter}
B.~Carter, \emph{Axisymmetric black hole has only two degrees of freedom},
  \href{https://doi.org/10.1103/PhysRevLett.26.331}{\emph{Phys. Rev. Lett.}
  {\bfseries 26} (1971) 331}.

\bibitem{Bekenstein_1972_1}
J.D.~Bekenstein, \emph{Transcendence of the law of baryon-number conservation
  in black-hole physics},
  \href{https://doi.org/10.1103/PhysRevLett.28.452}{\emph{Phys. Rev. Lett.}
  {\bfseries 28} (1972) 452}.

\bibitem{Bekenstein_1972_2}
J.D.~Bekenstein, \emph{Nonexistence of baryon number for static black holes},
  \href{https://doi.org/10.1103/PhysRevD.5.1239}{\emph{Phys. Rev. D} {\bfseries
  5} (1972) 1239}.

\bibitem{Bekenstein_1995}
J.D.~Bekenstein, \emph{Novel ``no-scalar-hair'' theorem for black holes},
  \href{https://doi.org/10.1103/PhysRevD.51.R6608}{\emph{Phys. Rev. D}
  {\bfseries 51} (1995) R6608}.

\bibitem{BEKENSTEIN1974}
J.D.~Bekenstein, \emph{Exact solutions of einstein-conformal scalar equations},
  \href{https://doi.org/https://doi.org/10.1016/0003-4916(74)90124-9}{\emph{Annals
  of Physics} {\bfseries 82} (1974) 535}.

\bibitem{BEKENSTEIN1975}
J.D.~Bekenstein, \emph{Black holes with scalar charge},
  \href{https://doi.org/https://doi.org/10.1016/0003-4916(75)90279-1}{\emph{Annals
  of Physics} {\bfseries 91} (1975) 75}.

\bibitem{BRONNIKOV1978_1}
K.~Bronnikov and Y.~Kireyev, \emph{Instability of black holes with scalar
  charge},
  \href{https://doi.org/https://doi.org/10.1016/0375-9601(78)90030-0}{\emph{Physics
  Letters A} {\bfseries 67} (1978) 95}.

\bibitem{Bronnikov1979_2}
K.A.~Bronnikov and A.V.~Khodunov, \emph{Scalar field and gravitational
  instability}, \href{https://doi.org/10.1007/BF00756667}{\emph{General
  Relativity and Gravitation} {\bfseries 11} (1979) 13}.

\bibitem{martinez_zaneli_1996}
C.~Mart\'{\i}nez and J.~Zanelli, \emph{Conformally dressed black hole in 2 + 1
  dimensions}, \href{https://doi.org/10.1103/PhysRevD.54.3830}{\emph{Phys. Rev.
  D} {\bfseries 54} (1996) 3830}.

\bibitem{Radu_Winstanley_2005}
E.~Radu and E.~Winstanley, \emph{Conformally coupled scalar solitons and black
  holes with negative cosmological constant},
  \href{https://doi.org/10.1103/PhysRevD.72.024017}{\emph{Phys. Rev. D}
  {\bfseries 72} (2005) 024017}.

\bibitem{Winstanley_2003}
E.~Winstanley, \emph{On the existence of conformally coupled scalar field hair
  for black holes in (anti-)de sitter space},
  \href{https://doi.org/10.1023/A:1022871809835}{\emph{Foundations of Physics}
  {\bfseries 33} (2003) 111}.

\bibitem{Winstanley_2005}
E.~Winstanley, \emph{Dressing a black hole with non-minimally coupled scalar
  field hair},
  \href{https://doi.org/10.1088/0264-9381/22/11/020}{\emph{Classical and
  Quantum Gravity} {\bfseries 22} (2005) 2233}.

\bibitem{MTZ}
C.~Mart\'{\i}nez, R.~Troncoso and J.~Zanelli, \emph{de sitter black hole with a
  conformally coupled scalar field in four dimensions},
  \href{https://doi.org/10.1103/PhysRevD.67.024008}{\emph{Phys. Rev. D}
  {\bfseries 67} (2003) 024008}.

\bibitem{MTZ2}
C.~Mart\'{\i}nez, R.~Troncoso and J.~Zanelli, \emph{Exact black hole solution
  with a minimally coupled scalar field},
  \href{https://doi.org/10.1103/PhysRevD.70.084035}{\emph{Phys. Rev. D}
  {\bfseries 70} (2004) 084035}.

\bibitem{martinez_troncoso_2006}
C.~Mart\'{\i}nez and R.~Troncoso, \emph{Electrically charged black hole with
  scalar hair}, \href{https://doi.org/10.1103/PhysRevD.74.064007}{\emph{Phys.
  Rev. D} {\bfseries 74} (2006) 064007}.

\bibitem{Poisson_1991}
E.~Poisson, \emph{Quadratic gravity as hair tonic for black holes},
  \href{https://doi.org/10.1088/0264-9381/8/4/009}{\emph{Classical and Quantum
  Gravity} {\bfseries 8} (1991) 639}.

\bibitem{Kanti_etal_1996}
P.~Kanti, N.E.~Mavromatos, J.~Rizos, K.~Tamvakis and E.~Winstanley,
  \emph{Dilatonic black holes in higher curvature string gravity},
  \href{https://doi.org/10.1103/PhysRevD.54.5049}{\emph{Phys. Rev. D}
  {\bfseries 54} (1996) 5049}.

\bibitem{Alexeyev_etal_2009}
S.~Alexeyev, A.~Barrau and K.A.~Rannu, \emph{Internal structure of a
  maxwell-gauss-bonnet black hole},
  \href{https://doi.org/10.1103/PhysRevD.79.067503}{\emph{Phys. Rev. D}
  {\bfseries 79} (2009) 067503}.

\bibitem{Yagi_etal_2012}
K.~Yagi, L.C.~Stein, N.~Yunes and T.~Tanaka, \emph{Post-newtonian,
  quasicircular binary inspirals in quadratic modified gravity},
  \href{https://doi.org/10.1103/PhysRevD.85.064022}{\emph{Phys. Rev. D}
  {\bfseries 85} (2012) 064022}.

\bibitem{Ayzenberg_etal_2014}
D.~Ayzenberg, K.~Yagi and N.~Yunes, \emph{Linear stability analysis of
  dynamical quadratic gravity},
  \href{https://doi.org/10.1103/PhysRevD.89.044023}{\emph{Phys. Rev. D}
  {\bfseries 89} (2014) 044023}.

\bibitem{Blazquez_Salcedo_etal_2016}
J.L.~Bl\'azquez-Salcedo, C.F.B.~Macedo, V.~Cardoso, V.~Ferrari, L.~Gualtieri,
  F.S.~Khoo et~al., \emph{Perturbed black holes in
  einstein-dilaton-gauss-bonnet gravity: Stability, ringdown, and
  gravitational-wave emission},
  \href{https://doi.org/10.1103/PhysRevD.94.104024}{\emph{Phys. Rev. D}
  {\bfseries 94} (2016) 104024}.

\bibitem{Bhattacharya_Chakraborty_2017}
S.~Bhattacharya and S.~Chakraborty, \emph{Constraining some horndeski gravity
  theories}, \href{https://doi.org/10.1103/PhysRevD.95.044037}{\emph{Phys. Rev.
  D} {\bfseries 95} (2017) 044037}.

\bibitem{Doneva_etal_2018}
D.D.~Doneva, S.~Kiorpelidi, P.G.~Nedkova, E.~Papantonopoulos and
  S.S.~Yazadjiev, \emph{Charged gauss-bonnet black holes with curvature induced
  scalarization in the extended scalar-tensor theories},
  \href{https://doi.org/10.1103/PhysRevD.98.104056}{\emph{Phys. Rev. D}
  {\bfseries 98} (2018) 104056}.

\bibitem{Tattersall_etal_2018}
O.J.~Tattersall, P.G.~Ferreira and M.~Lagos, \emph{Speed of gravitational waves
  and black hole hair},
  \href{https://doi.org/10.1103/PhysRevD.97.084005}{\emph{Phys. Rev. D}
  {\bfseries 97} (2018) 084005}.

\bibitem{Brihaye_Hartmann_2018}
Y.~Brihaye and B.~Hartmann, \emph{Critical phenomena of charged
  einstein–gauss–bonnet black holes with charged scalar hair},
  \href{https://doi.org/10.1088/1361-6382/aad389}{\emph{Classical and Quantum
  Gravity} {\bfseries 35} (2018) 175008}.

\bibitem{Brihaye_Hartmann_2019}
Y.~Brihaye and B.~Hartmann, \emph{Spontaneous scalarization of charged black
  holes at the approach to extremality},
  \href{https://doi.org/https://doi.org/10.1016/j.physletb.2019.03.043}{\emph{Physics
  Letters B} {\bfseries 792} (2019) 244}.

\bibitem{BRIHAYE_etal_2019}
Y.~Brihaye, C.~Herdeiro and E.~Radu, \emph{The scalarised schwarzschild-nut
  spacetime},
  \href{https://doi.org/https://doi.org/10.1016/j.physletb.2018.11.022}{\emph{Physics
  Letters B} {\bfseries 788} (2019) 295}.

\bibitem{BumHoon_etal_2019}
B.-H.~Lee, W.~Lee and D.~Ro, \emph{Expanded evasion of the black hole no-hair
  theorem in dilatonic einstein-gauss-bonnet theory},
  \href{https://doi.org/10.1103/PhysRevD.99.024002}{\emph{Phys. Rev. D}
  {\bfseries 99} (2019) 024002}.

\bibitem{CAMPBELL1990_1}
B.A.~Campbell, M.~Duncan, N.~Kaloper and K.A.~Olive, \emph{Axion hair for kerr
  black holes},
  \href{https://doi.org/https://doi.org/10.1016/0370-2693(90)90227-W}{\emph{Physics
  Letters B} {\bfseries 251} (1990) 34}.

\bibitem{CAMPBELL1991_dyonbh}
B.A.~Campbell, N.~Kaloper and K.A.~Olive, \emph{Axion hair for dyon black
  holes},
  \href{https://doi.org/https://doi.org/10.1016/0370-2693(91)90474-5}{\emph{Physics
  Letters B} {\bfseries 263} (1991) 364}.

\bibitem{Kimyeong_Weinberg_1991_axion}
K.~Lee and E.J.~Weinberg, \emph{Charged black holes with scalar hair},
  \href{https://doi.org/10.1103/PhysRevD.44.3159}{\emph{Phys. Rev. D}
  {\bfseries 44} (1991) 3159}.

\bibitem{Bardoux_2012}
Y.~Bardoux, M.M.~Caldarelli and C.~Charmousis, \emph{Conformally coupled scalar
  black holes admit a flat horizon due to axionic charge},
  \href{https://doi.org/10.1007/JHEP09(2012)008}{\emph{Journal of High Energy
  Physics} {\bfseries 2012} (2012) 8}.

\bibitem{scalar1_PhysRevLett.121.101102}
C.A.R.~Herdeiro, E.~Radu, N.~Sanchis-Gual and J.A.~Font, \emph{Spontaneous
  scalarization of charged black holes},
  \href{https://doi.org/10.1103/PhysRevLett.121.101102}{\emph{Phys. Rev. Lett.}
  {\bfseries 121} (2018) 101102}.

\bibitem{Fernandes_2019}
P.G.S.~Fernandes, C.A.R.~Herdeiro, A.M.~Pombo, E.~Radu and N.~Sanchis-Gual,
  \emph{Spontaneous scalarisation of charged black holes: coupling dependence
  and dynamical features},
  \href{https://doi.org/10.1088/1361-6382/ab23a1}{\emph{Classical and Quantum
  Gravity} {\bfseries 36} (2019) 134002}.

\bibitem{Myung2019}
Y.S.~Myung and D.-C.~Zou, \emph{Stability of scalarized charged black holes in
  the einstein--maxwell--scalar theory},
  \href{https://doi.org/10.1140/epjc/s10052-019-7176-7}{\emph{The European
  Physical Journal C} {\bfseries 79} (2019) 641}.

\bibitem{Jhep_main2019}
D.~Astefanesei, C.~Herdeiro, A.~Pombo and E.~Radu,
  \emph{Einstein-maxwell-scalar black holes: classes of solutions, dyons and
  extremality}, \href{https://doi.org/10.1007/JHEP10(2019)078}{\emph{Journal of
  High Energy Physics} {\bfseries 2019} (2019) 78}.

\bibitem{scalar2_Fernandes_2019}
P.G.S.~Fernandes, C.A.R.~Herdeiro, A.M.~Pombo, E.~Radu and N.~Sanchis-Gual,
  \emph{Spontaneous scalarisation of charged black holes: coupling dependence
  and dynamical features},
  \href{https://doi.org/10.1088/1361-6382/ab23a1}{\emph{Classical and Quantum
  Gravity} {\bfseries 36} (2019) 134002}.

\bibitem{scalar3}
J.L.~Blázquez-Salcedo, C.A.~Herdeiro, J.~Kunz, A.M.~Pombo and E.~Radu,
  \emph{Einstein-maxwell-scalar black holes: The hot, the cold and the bald},
  \href{https://doi.org/https://doi.org/10.1016/j.physletb.2020.135493}{\emph{Physics
  Letters B} {\bfseries 806} (2020) 135493}.

\bibitem{scalar4_PhysRevD}
R.A.~Konoplya and A.~Zhidenko, \emph{Analytical representation for metrics of
  scalarized einstein-maxwell black holes and their shadows},
  \href{https://doi.org/10.1103/PhysRevD.100.044015}{\emph{Phys. Rev. D}
  {\bfseries 100} (2019) 044015}.

\bibitem{massive_scalar_PhysRevD.100.124055}
D.-C.~Zou and Y.S.~Myung, \emph{Scalarized charged black holes with scalar mass
  term}, \href{https://doi.org/10.1103/PhysRevD.100.124055}{\emph{Phys. Rev. D}
  {\bfseries 100} (2019) 124055}.

\bibitem{massive_self_int_scalar}
P.G.~Fernandes, \emph{Einstein–maxwell-scalar black holes with massive and
  self-interacting scalar hair},
  \href{https://doi.org/https://doi.org/10.1016/j.dark.2020.100716}{\emph{Physics
  of the Dark Universe} {\bfseries 30} (2020) 100716}.

\bibitem{Promsirietal}
C.~Promsiri, T.~Tangphati, E.~Hirunsirisawat and S.~Ponglertsakul,
  \emph{Scalarization of planar anti--de sitter charged black holes in
  einstein-maxwell-scalar theory},
  \href{https://doi.org/10.1103/PhysRevD.108.024015}{\emph{Phys. Rev. D}
  {\bfseries 108} (2023) 024015}.

\bibitem{Mayo_Bekenstein_1996}
A.E.~Mayo and J.D.~Bekenstein, \emph{No hair for spherical black holes: Charged
  and nonminimally coupled scalar field with self-interaction},
  \href{https://doi.org/10.1103/PhysRevD.54.5059}{\emph{Phys. Rev. D}
  {\bfseries 54} (1996) 5059}.

\bibitem{Hong_suzuki_yamada_PLB}
J.-P.~Hong, M.~Suzuki and M.~Yamada, \emph{Charged black holes in non-linear
  q-clouds with o(3) symmetry},
  \href{https://doi.org/https://doi.org/10.1016/j.physletb.2020.135324}{\emph{Physics
  Letters B} {\bfseries 803} (2020) 135324}.

\bibitem{Hong_suzuki_yamada_PRL}
J.-P.~Hong, M.~Suzuki and M.~Yamada, \emph{Spherically symmetric scalar hair
  for charged black holes},
  \href{https://doi.org/10.1103/PhysRevLett.125.111104}{\emph{Phys. Rev. Lett.}
  {\bfseries 125} (2020) 111104}.

\bibitem{Herdeiro_radu_2020}
C.A.R.~Herdeiro and E.~Radu, \emph{Spherical electro-vacuum black holes with
  resonant, scalar q-hair},
  \href{https://doi.org/10.1140/epjc/s10052-020-7976-9}{\emph{The European
  Physical Journal C} {\bfseries 80} (2020) 390}.

\bibitem{CS1}
S.~Chandrasekhar, \emph{Dynamical instability of gaseous masses approaching the
  schwarzschild limit in general relativity},
  \href{https://doi.org/10.1103/PhysRevLett.12.114}{\emph{Phys. Rev. Lett.}
  {\bfseries 12} (1964) 114}.

\bibitem{CS2}
S.~Chandrasekhar, \emph{Dynamical instability of gaseous masses approaching the
  schwarzschild limit in general relativity},
  \href{https://doi.org/10.1103/PhysRevLett.12.114}{\emph{Phys. Rev. Lett.}
  {\bfseries 12} (1964) 114}.

\bibitem{CS3}
S.~{Chandrasekhar} and R.F.~{Tooper}, \emph{The dynamical instability of the
  white-dwarf configurations approaching the limiting mass.},
  \href{https://doi.org/10.1086/147883}{\emph{Astrophysical Journal} {\bfseries
  139} (1964) 1396}.

\bibitem{Bardeen_1966}
J.M.~{Bardeen}, K.S.~{Thorne} and D.W.~{Meltzer}, \emph{A catalogue of methods
  for studying the normal modes of radial pulsation of general-relativistic
  stellar models}, \href{https://doi.org/10.1086/148791}{\emph{Astrophysical
  Journal} {\bfseries 145} (1966) 505}.

\bibitem{Hawking_Ellis_1973}
S.W.~Hawking and G.F.R.~Ellis, \emph{The Large Scale Structure of Space-Time},
  Cambridge Monographs on Mathematical Physics, Cambridge University Press
  (1973).

\bibitem{Bekenstein1973}
J.D.~Bekenstein, \emph{Extraction of energy and charge from a black hole},
  \href{https://doi.org/10.1103/PhysRevD.7.949}{\emph{Phys. Rev. D} {\bfseries
  7} (1973) 949}.

\bibitem{HOD3}
S.~Hod, \emph{Stability of the extremal reissner–nordström black hole to
  charged scalar perturbations},
  \href{https://doi.org/https://doi.org/10.1016/j.physletb.2012.06.043}{\emph{Physics
  Letters B} {\bfseries 713} (2012) 505}.

\bibitem{HOD1}
S.~Hod, \emph{No-bomb theorem for charged reissner–nordström black holes},
  \href{https://doi.org/https://doi.org/10.1016/j.physletb.2012.12.013}{\emph{Physics
  Letters B} {\bfseries 718} (2013) 1489}.

\end{thebibliography}
\end{document}